\documentclass[epj]{svjour} 
\usepackage[numbers,sort&compress]{natbib}
\usepackage{times}
\usepackage{graphicx}
\usepackage{amssymb}

\newcommand{\affiliation}[1]{\inst{#1}}
\newcommand{\ISKP}{1}
\newcommand{\HH}{2}
\newcommand{\IKP}{3}
\newcommand{\AISKP}{Helmholtz-Institut f\"ur Strahlen- und Kernphysik,
Universit\"at Bonn, D-53115 Bonn, Germany}
\newcommand{\AHH}{Institut f\"ur Experimentalphysik, Universit\"at Hamburg,
D-22761 Hamburg, Germany}
\newcommand{\AIKP}{Institut f\"ur Kernphysik, Forschungszentrum J\"ulich,
D-52425 J\"ulich, Germany}

\newcommand{\authors}{
	\author{
{D.~Albers} \affiliation{\HH} 
\and{F.~Bauer} \affiliation{\HH}
\and{J.~Bisplinghoff} \affiliation{\ISKP} 
\and{R.~Bollmann} \affiliation{\HH} 
\and{K.~B\"u\ss{}er} \affiliation{\HH} 
\and{M.~Busch} \affiliation{\ISKP} 
\and{R.~Daniel} \affiliation{\ISKP} 
\and{O.~Diehl} \affiliation{\ISKP} 
\and{F.~Dohrmann} \affiliation{\HH}
\and{H.P.~Engelhardt} \affiliation{\ISKP}
\and{J.~Ernst} \affiliation{\ISKP}
\and{P.D.~Eversheim} \affiliation{\ISKP} 
\and{M.~Gasthuber} \affiliation{\HH}
\and{R.~Gebel} \affiliation{\IKP} 
\and{J.~Greiff} \affiliation{\HH} 
\and{A.~Gro\ss} \affiliation{\HH} 
\and{R.~Gro\ss-Hardt} \affiliation{\ISKP}  
\and{S.~Heider} \affiliation{\ISKP}
\and{A.~Heine} \affiliation{\ISKP}
\and{F.~Hinterberger} \affiliation{\ISKP}
\and{T.~H\"uskes} \affiliation{\ISKP} 
\and{M.~Igelbrink} \affiliation{\HH}
\and{M.~Jeske} \affiliation{\ISKP}
\and{R.~Langkau} \affiliation{\HH}
\and{J.~Lindlein} \affiliation{\HH}
\and{R.~Maier} \affiliation{\IKP} 
\and{R.~Maschuw} \affiliation{\ISKP}
\and{F.~Mosel} \affiliation{\ISKP}
\and{D.~Prasuhn} \affiliation{\IKP}
\and{H.~Rohdje\ss}\affiliation{\ISKP}
\and{D.~Rosendaal} \affiliation{\ISKP}
\and{P.~von~Rossen} \affiliation{\IKP}
\and{N.~Scheid} \affiliation{\ISKP}
\and{N.~Schirm} \affiliation{\HH}
\and{M.~Schulz-Rojahn} \affiliation{\ISKP}
\and{F.~Schwandt} \affiliation{\ISKP}
\and{V.~Schwarz} \affiliation{\ISKP}
\and{W.~Scobel} \affiliation{\HH} 
\and{S.~Thomas} \affiliation{\ISKP}
\and{H.-J.~Trelle} \affiliation{\ISKP}
\and{E.~Weise} \affiliation{\ISKP}
\and{A.~Wellinghausen} \affiliation{\HH}
\and{K.~Woller} \affiliation{\HH}
\and{R.~Ziegler} \affiliation{\ISKP}
}
} 

\newcommand{\Beginabstract}{\abstract}
\newcommand{\Endabstract}{}
\newcommand{\collaboration}[1]{}
\newcommand{\mypacs}{
\PACS{
	{25.40.Cm}{Elastic proton scattering} \and
	{13.75.Cs}{Nucleon-nucleon interactions} \and
	{13.85.Dz}{Elastic scattering } \and
	{21.30.-x}{Nuclear forces}
}
}

\newcommand{\institutes}{
\institute{\AISKP} \and 
\institute{\AHH} \and 
\institute{\AIKP} }
\renewcommand{\institutes}{
\institute{\AISKP \and \AHH \and \AIKP}
\offprints{H.~Rohdje\ss, HISKP, Universit\"at Bonn, Nussallee 14-16,
	D-53115 Bonn, Germany}
\mail{rohdjess@iskp.uni-bonn.de}}
\newcommand{\preprint}[1]{}
\newcommand{\printfigures}{}

\newcommand{\Beginack}{\begin{acknowledgement}}
\newcommand{\Endack}{\end{acknowledgement}}
\newcommand{\figwidth}{8.5cm}
\newcommand{\widefigwidth}{18cm}
\renewcommand{\figwidth}{\columnwidth}
\renewcommand{\widefigwidth}{\textwidth}

\newcommand{\SidecaptionBeg}{\sidecaption}
\newcommand{\SidecaptionEnd}{}

\newcommand{\begindescription}[1]{\begin{description}[#1]}

\newcommand{\MacroRev}{$Revision: 1.26 $}

\newcommand{\marm}{\rm}
\newcommand{\mylabel}[1]{\label{#1}
       \makebox[0cm][r]{\raisebox{5mm}[0mm][0mm]{\framebox{\tt #1}}}}
\newcommand{\mylabelnr}[1]{\label{#1}
       \makebox[0cm][r]{\framebox{\tt #1}\ \ \ }}
\renewcommand{\mylabel}[1]{\label{#1}}
\renewcommand{\mylabelnr}[1]{\label{#1}}

\newcommand{\cms}{c.m.}
\newcommand{\kd}{\ensuremath{\alpha}}
\newcommand{\kdc}{\ensuremath{\alpha_{\rm cut}}}
\newcommand{\kdmin}{\ensuremath{\alpha_{\rm min}}}
\newcommand{\kdmax}{\ensuremath{\alpha_{\rm max}}}

\newcommand{\Tp}{\ensuremath{T_p}}
\newcommand{\Sec}[1]{Sect.~\ref{#1}}

\newcommand{\fhf}{\ensuremath{\nu_{\rm RF}}}
\newcommand{\mom}{\ensuremath{p}}
\newcommand{\cm}{\mbox{\,cm\ }}
\newcommand{\evspace}{\ensuremath \:\:\!\!\!}

\newcommand{\MeV}{\ensuremath{\mathrm {\,Me}{\evspace}V}}
\newcommand{\GeVc}{\ensuremath{\mathrm {\,Ge}{\evspace}V/c}}
\newcommand{\MeVc}{\ensuremath{\mathrm{\,Me}{\evspace}V/c}}

\newcommand{\GeV}{\ensuremath{\mathrm{\,Ge}{\evspace}V}}

\newcommand{\mytheta}{\theta}
\newcommand{\bcm}{\ensuremath{\beta_{\rm \cms}}}
\newcommand{\gcm}{\ensuremath{\gamma_{\rm \cms}}}
\newcommand{\myphi}{\ensuremath{\phi}}
\newcommand{\Ph}[1]{\ensuremath{\myphi_{#1}}}
\newcommand{\Thlab}{\ensuremath{\mytheta_{\rm lab}}\ }
\newcommand{\Thlabi}[1]{\ensuremath{\mytheta_{\rm lab,#1}}}
\newcommand{\Thlabx}{\ensuremath{\mytheta_{\rm lab}}}
\newcommand{\Thcms}{\ensuremath{\mytheta_{\rm \cms}}\ }
\newcommand{\Thcmsx}{\ensuremath{\mytheta_{\rm \cms}}}
\newcommand{\Thcmsi}[1]{\ensuremath{\mytheta_{\rm \cms,#1}}}

\newcommand{\sx}{\mbox{\,s}}
\newcommand{\s}{\mbox{\,s\ }}
\newcommand{\msx}{\mbox{\,ms}}
\newcommand{\ms}{\mbox{\,ms\ }}
\newcommand{\nsx}{\mbox{\,ns}}

\newcommand{\cmx}{\mbox{\,cm}}
\newcommand{\mmx}{\mbox{\,mm}}

\newcommand{\perc}{\mbox{\,\%\ }}
\newcommand{\percx}{\mbox{\,\%}}

\newcommand{\kHzx}{\mbox{\,kHz}}

\newcommand{\MHz}{\mbox{\,MHz\ }}
\newcommand{\grad}{\ensuremath{^{\circ}}\ }
\newcommand{\gradx}{\ensuremath{^{\circ}}}
\newcommand{\mugx}{\mbox{\,$\mu$g}}

\newcommand{\Arg}{\ensuremath{(\mom,\Thcmsx)}}
\newcommand{\Npp}[2]{\ensuremath{N_{\rm #1}^{\rm #2}}}
\newcommand{\NCH}{\Npp{CH_2}{}}
\newcommand{\Nppel}{\Npp{pp}{el.}}
\newcommand{\Nppin}{\Npp{pp}{in.}}
\newcommand{\NC}{\Npp{C}{}}
\newcommand{\NCa}[1]{\Npp{C}{(#1)}}
\newcommand{\CH}{\ensuremath{\rm CH_2}}
\newcommand{\C}{\ensuremath{\rm C}}
\newcommand{\Nch}{\ensuremath{N_{\CH}}}
\newcommand{\Nc}{\ensuremath{N_{\C}}}
\newcommand{\fnorm}{\ensuremath{{\cal L}}}

\newcommand{\EM}[1]{{\em #1}}

\newcommand{\absolute}{\EM{absolute} }
\newcommand{\DTZ}{\ensuremath{\Delta_{Tz}}}
\newcommand{\Dphi}{\ensuremath{\Delta\myphi}}
\newcommand{\DE}[1]{\ensuremath{\Delta E_{#1}}}
\newcommand{\DEe}{\ensuremath{\DE{1}}}
\newcommand{\DEz}{\ensuremath{\DE{2}}}
\newcommand{\DEez}{\ensuremath{\DE{1(2)}}}

\newcommand{\dsdt}[1]{\ensuremath{{\marm \frac{d\sigma_{#1}}{dt}}}}

\newcommand{\dsdoc}[1]{\ensuremath{{\marm d\sigma_{#1}/d\Omega_{\cms}}}}

\newcommand{\dsdocd}[1]{\ensuremath{{\marm \frac{d\sigma_{#1}}{d\Omega_{\cms}}}}}
\newcommand{\chiz}{\ensuremath{{\chi^2}}}

\newcommand{\Fig}[1]{Fig.~\ref{#1}}
\newcommand{\Figs}[1]{Figs.~\ref{#1}}

\newcommand{\Eq}[1]{Eq.~(\ref{#1})}
\newcommand{\Eqs}[2]{Eqs.~(\ref{#1}) and (\ref{#2})}

\newcommand{\REL}{\ensuremath{\left|\left.\frac{{\rm d}E}{{\rm d}z}\right|_{T<T_{\rm cut}}\right|}}
\newcommand{\RELp}{\ensuremath{\left|\left.\frac{{\rm d}E}{{\rm d}z}(p)\right|_{T<T_{\rm cut}}\right|}}
\newcommand{\RELN}{\ensuremath{\left.\frac{{\rm d}E}{{\rm d}z}\right|_{T<T_{\rm cut}}}}
\newcommand{\LSEM}{\ensuremath{L_{\rm SEM}}}
\newcommand{\kSEM}{\ensuremath{k_{\rm SEM}}}
\newcommand{\ISEM}{\ensuremath{I_{\rm SEM}}}
\newcommand{\LPIN}{\ensuremath{L_{\rm PIN}}}

\newcommand{\pref}{\ensuremath{p_{\rm ref}}}

\newcommand{\Pel}{\ensuremath{P_{\rm el.}}}
\newcommand{\Probp}{\ensuremath{P^{(g)}}}
\newcommand{\Probpa}{\ensuremath{\Probp_{(a)}}}
\newcommand{\Probpb}{\ensuremath{\Probp_{(b)}}}
\newcommand{\Prob}[1]{\ensuremath{\Probp(#1)}}
\newcommand{\Proba}[1]{\ensuremath{\Probpa(#1)}}
\newcommand{\Probb}[1]{\ensuremath{\Probpb(#1)}}
\newcommand{\Pcut}{\ensuremath{P_{\rm cut}}}

\newcommand{\For}{\ensuremath{\mathrm{for}}}

\newcommand{\vecQ}{\ensuremath{\vec{q}}}
\newcommand{\vecQa}{\ensuremath{\vec{q}_{(a)}}}
\newcommand{\vecQb}{\ensuremath{\vec{q}_{(b)}}}

\newcommand{\dE}{\ensuremath{{\rm d}E}}
\newcommand{\dx}{\ensuremath{{\rm d}x}}
\newcommand{\dEdx}{\ensuremath{\frac{\dE}{\dx}}}

\newcommand{\pp}{\ensuremath{\rm pp}}

\newcommand{\Nor}[1]{\ensuremath{{\cal N}_{#1}}}
\newcommand{\Noi}{\Nor{i}}

\newcommand{\myrho}{\rho}

\newcommand{\Gy}{{\rm Gy}}

\newcommand{\dd}[1]{\ensuremath{{\rm d}#1}}

\newcommand{\dt}{\dd{t}}

\newcommand{\dD}{\dd{D}}

\newcommand{\Lum}{\ensuremath{\mathcal{L}}}

\newcommand{\hden}{\ensuremath{\myrho}}
\newcommand{\ahden}{\ensuremath{\overline{\myrho}}}
\newcommand{\hdenn}{\ensuremath{\hden_0}}
\newcommand{\rhden}{\ensuremath{\frac{\ahden}{\hdenn }}}
\newcommand{\rhdenn}{\ensuremath{\frac{\hden}{\hdenn }}}
\newcommand{\vrhden}{\ensuremath{\ahden/\hdenn}}
\newcommand{\vrhdenn}{\ensuremath{\hden/\hdenn}}

\newcommand{\vrhdennt}{\ensuremath{\hden(t)/\hdenn}}

\newcommand{\Star}{\protect\includegraphics[width=2mm]{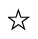}}
\newcommand{\Cross}{\protect\includegraphics[width=2mm]{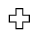}}
\newcommand{\Ast}{\protect\includegraphics[width=2mm]{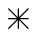}}
\newcommand{\Circ}{\ensuremath{\circ}}

\newcommand{\siehe}{see}
\newcommand{\ttfon}{}

\newcommand{\Revision}{$Revision: 1.53 $}
\newcommand{\Thisrevision}{\Revision, Macros: \MacroRev}
\begin{document}
\preprint{\Thisrevision}
\title{A Precision Measurement of pp Elastic Scattering Cross
Sections at Intermediate Energies}
 
\authors

\institutes

\collaboration{EDDA Collaboration}

\date{\today}
 
\Beginabstract{
We have measured differential cross sections for \pp\ elastic scattering with 
internal fiber targets in the recirculating beam of the proton synchrotron COSY.
Measurements were made continuously during acceleration
for projectile kinetic energies between 0.23 and 2.59~\GeV\ in 
the angular range $30\grad \leq \Thcmsx \leq 90\grad$. Details of the 
apparatus and the data analysis are given and the resulting excitation 
functions and angular distributions presented. The precision of each
data point is typically better than 4\%, and a relative normalization
uncertainty of only 2.5\% within an excitation function has been reached.
The impact on phase shift analysis as well as upper bounds on possible
resonant contributions in lower partial waves are discussed.
}\Endabstract

\mypacs

\maketitle

%
%
\section{Introduction}\mylabel{introduction}
%
Elastic nucleon-nucleon scattering is a process fundamental to
understanding nuclear forces. Its precise experimental
knowledge bears on such basic questions as the confinement of
quarks, the limits to the
validity of meson exchange models in the 
nuclear regime and a basis upon which to test QCD inspired models
at intermediate energies. Its knowledge also forms the basis of a broad range
of applications in nuclear and heavy ion physics, e.g. as ingredients
to models of reaction dynamics, excited nuclear matter and transport phenomena.
Consequently, many experimental and theoretical studies
(see \cite{Lechanoine:1993} and references therein) have been devoted
to this subject. The data base has about doubled over
the past decade \cite{Haeberli:1997,vonPrzewoski:1998ye,Rathmann:1998,Ball:2000}, and global phase shifts
- a convenient tool of parameterizing existing
experimental knowledge - are on a sound basis up to about
1~GeV in kinetic beam energy
\cite{arndt94d,Arndt:1997if,Arndt:2000xc}.

Closer inspection shows, however, that the vast majority of the
data is below 0.8~GeV of kinetic beam energy. Above that beam
energy, the data base used to be increasingly sparse 
and somewhat uncertain in normalization,
hampering both reliable phase shift analysis and conclusions
towards the physics of strong interaction.

The EDDA experiment was designed to provide \pp\ elastic
scattering data up to 2.5~GeV - precise and consistent in 
normalization - for all of the above purposes.
In particular, it was meant to supply data for
phase shift analysis up to
that
projectile energy, and to test claims \cite{ball94}
and predictions, e.g. \cite{Mulders:1978yi,Mulders:1980vx,LaFrance:1986pc,Gonzalez:1987gj}, of dibaryonic resonances. To this end 
EDDA measured
spin-averaged differential cross sections \cite{albers97}
as well as transverse analyzing
powers \cite{Altmeier:2000} and spin correlation coefficients
\cite{Bauer:2002zm}, utilizing the proton beams available at
the cooler synchrotron COSY \cite{Maier:1997} at FZ J\"ulich.

An account of the results has already been
given in \cite{albers97}, based on roughly 40\% of the
data available now. The purpose of this
paper is to describe experimental
details and to present a more
elaborate analysis.
In particular, corrections for radiation
damage of the target was not included in the analysis of
Ref. \cite{albers97} which lead to cross
sections systematically smaller by $\approx 3$\% for momenta above
about 2~GeV/c. In addition, a refined separation of proton-carbon
scattering events was accomplished by pattern recognition techniques,
resulting in reduced statistical errors. In total the statistical
precision was improved by up to a factor of 3. 

The experimental setup   
 is described in \Sec{exp}, 
while \Sec{offline} focuses on a detailed description of the data analysis
including  
event selection, vertex and angle reconstruction, background subtraction and 
normalization. \Sec{results} presents angular distributions and excitation functions,
\Sec{disc} discusses their impact on phase shift analysis and
 hypothetical dibaryonic resonances.  
%
%
%
\section{The EDDA experiment}\mylabel{exp}
%
EDDA was conceived as an experiment using the {\em internal} beam of the
cooler synchrotron COSY \cite{Maier:1997}, so as 
{\em to measure continuous excitation functions
during beam acceleration}. It is placed at a beam waist in one
of the straight sections of COSY's race-track shaped lattice.
With the protons recirculating with about 1.0-1.5~MHz,
inherently thin, polarized atomic beam targets
for the measurement of spin observables can be utilized with
sufficient luminosity. However, for cross section measurements as
reported here, thin
\CH--fiber targets have been used, as they allow for a continuous
monitoring of the instantaneous luminosity.
%
\subsection{Targets and Detector}\mylabel{dettar}
%
\CH--fibers were strung horizontally between the prongs of a fork,
which could be moved vertically by a magnet-driven linear actuator to
put the fiber into (and 
out of) the COSY beam. The thickness of the fibers was chosen
as a compromise between long beam-lifetimes and sufficient sturdiness  
when exposed to the beam.
A cross-section of $4\times 5$~$\mu{\rm m}^2$ \CH\ proved to be a good
choice, allowing a few $10^8$ protons recirculating in the
COSY ring at about 1.5~MHz without target-failure.
Target fibers were prepared using a microtome to cut 4~$\mu$m
fibers from a commercially available polypropylene (\CH) foil
of 5~$\mu$m thickness.
To prevent build-up of charges due to ionization by the COSY-beam,
the surface was made conducting by a thin (20~$\mu$g/cm$^2$) aluminum
coating, such that the target was grounded through the supporting fork
made from aluminum.

The background resulting from the carbon content (and the Al coating)
needs to be subtracted offline; It  was measured using 5~$\mu$m thick
carbon fibers suitably coated with Al. 
\CH--targets lose {\em slowly} hydrogen content upon beam exposure.
Therefore, a reservoir of targets was kept inside
the COSY vacuum to ensure essentially uninterrupted 
operation of EDDA during data taking.
In first order, the relative normalization of the excitation functions
is not affected by the slow hydrogen loss since the time period of
about 2~s for measuring a complete excitation function during beam
acceleration is small with respect to the hydrogen loss rate. 

The detector as used for cross section measurements is
shown schematically in \Fig{fig:detector}: 
A scintillator hodoscope surrounds the beam pipe downstream the 
target 
covering the angular range $10\gradx \le \Thlabx \le 72\gradx$ and
subtending about 85\perc of $4\pi$ in the \cms\ for proton-proton elastic scattering. 
It consists of scintillator bars and rings designed to
measure the points, at which the scattered proton
and its recoil partner traverse the hodoscope
in terms of azimuthal and polar angle ($\varphi_{1,2}$ and $\Thlabi{1,2}$,
respectively). The kinematic relationships between
these angles (see \Sec{cuts}) are used to identify elastic scattering events.

The inner hodoscope layer consists of 32 scintillator bars 
made of BC408 and read out on both ends via lucite light guides with Hamamatsu 
R1355 photomultipliers (PM) \cite{Ackerstaff:1993vy}. The scintillator
bars have a triangular cross 
section (\Fig{fig:elements}~(a)).
Their overlap ensures that each charged particle originating at the
target deposits energy in two adjacent bars when
traversing the layer.
The fractional light output from adjacent scintillator bars is used in the
offline analysis (see \Sec{vertex}) to give the azimuthal angle to a
precision about five times better than would be
possible on the basis of granularity alone \cite{bisplinghoff93}.


\begin{figure}
\begin{center}
\includegraphics[width=\figwidth]{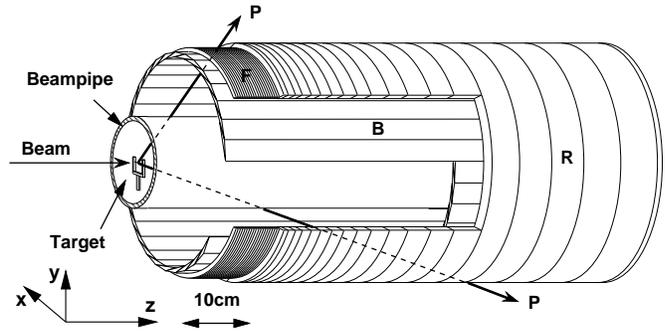}
\end{center}
\caption{ The EDDA detector (not to scale):
          target: fiber (\CH\ or C); 
          B: scintillator bars; R: scintillator semi-rings; 
          F: semi-rings made of scintillating fibers.
	}
\mylabel{fig:detector}
\end{figure}



\begin{figure}
\begin{center}
\includegraphics[width=\figwidth]{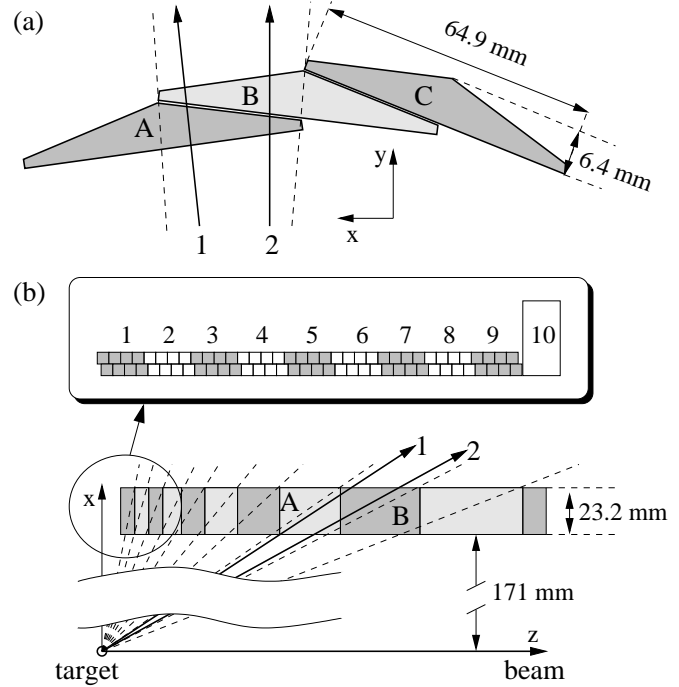}
\end{center}
\caption{ Cross sectional view of adjacent scintillator 
          bars (a) and semi-rings (b), crossed by two proton
trajectories which can be distinguished by the fractional light output
of elements A and B. The first 9 semi-rings are made up of 
scintillating fibers as shown in the figure.}
\mylabel{fig:elements}           
\end{figure}


The outer layer is composed of scintillator rings for 
angles $\Thlabx \le 52\gradx$, split into left and right semi-rings to
allow radial readout of the scintillation light in direction
$\pm y$. After total 
reflection by 90\grad\ to the top or bottom of 
the detector, the light is collected via light-guides by Hamamatsu
R1450 photomultipliers. 
The width of the rings varies along the beam
(cf. \Fig{fig:elements}~(b)) such that each proton trajectory crosses
2, in some cases 3 rings. Each ring covers an interval 
$\Delta\Thcmsx \approx 5\gradx$ in the center-of-mass frame for elastic 
\pp\ scattering. 
Again the ratio of light outputs from adjacent rings is used to improve the
resolution in (polar) angle over granularity.
The method becomes less and less effective with increasing
polar angle. Therefore, the scintillator rings
were replaced by a double layer of ($2\times 2$~mm$^2$ quadratic)
scintillating fibers for angles $\Thlabx \ge 52\gradx$ as shown in the
inset of \Fig{fig:elements}~(b) \cite{albers96b}.
The latter cover about 10~cm of the inner
layer close to the target, and they give -- by their granularity -- about the
same angular resolution as is obtained from the rings using
fractional light output analysis.

\subsection{Luminosity Monitors}\mylabel{lumiSetup}
%
\begin{figure}
\begin{center}
\includegraphics[width=\figwidth]{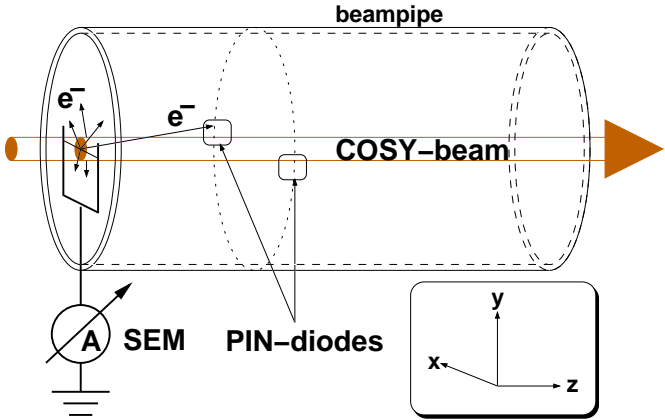}
\end{center}
\caption{Schematic setup of luminosity monitors.}
\mylabel{fig:setuplumi}
\end{figure}
%
Two independent, concurrent methods were applied to monitor luminosity
as sketched in \Fig{fig:setuplumi}. 
Both monitors detect electrons emerging from the fiber
target due to the beam-target interaction. The secondary electron monitor
(SEM) consists  
of a fast amperemeter \cite{scheid94} measuring continuously the current of electrons from 
ground to target, replacing low--energy secondary electrons emanating
from the target  
surface. High-energy $\delta$-electrons from elastic proton-electron 
scattering were detected 
in two PIN diodes located at $\Thlab \approx 40\gradx$ behind 
thin (250\mugx/cm$^2$) aluminum windows in small pockets of the
beam pipe.
Both rates scale
differently  
with projectile momentum and must be corrected for this dependence
when monitoring 
the luminosity consistently over the whole momentum range, see \Sec{lumi}.
%
\subsection{Cyclic operation}\mylabel{cycle}
%
Data are collected during synchrotron acceleration such that a
complete excitation function is measured in each acceleration
cycle. A typical cycle is shown in  
\Fig{fig:cycle}. 
\begin{figure}
\begin{center}
\includegraphics[width=\figwidth,bb=5 15 515 460]{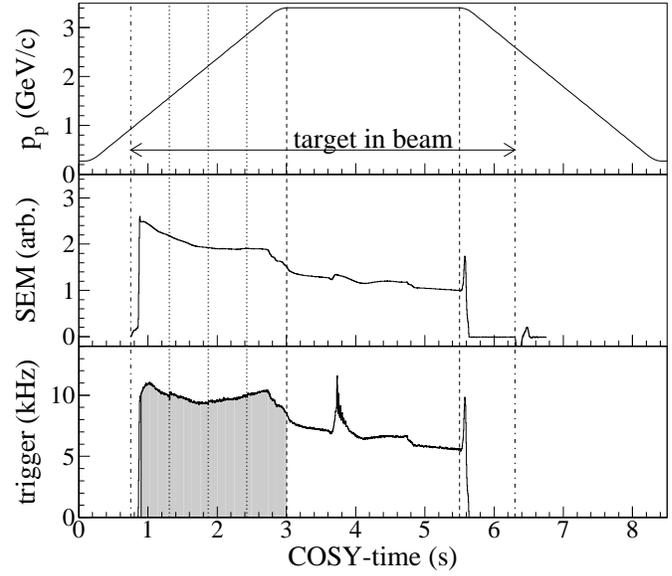}
\end{center}
\caption{ 
Timing during an experimental cycle. Shown is the projectile
momentum \mom\ (top), signal of the
SEM luminosity monitor (middle) and trigger rate (bottom) as a
function of cycle time. 
}\mylabel{fig:cycle}
\end{figure}
After injection of some 10$^7$ protons with $\mom$ = 275\MeVc, 
acceleration starts with a momentum ramp of 1.15\GeVc\ per second. At
$\mom \approx$ 0.7~\GeVc, the target is moved vertically
into the beam. The target remains in the beam until the maximum beam
momentum of $\mom \approx$ 3.3-3.4\GeVc\  
is reached (flattop). 
Since the horizontal beam position is not constant during
acceleration the beam was steered across the fiber in the flattop
(cf. \Fig{fig:vertex}), such that any effect relating to the position
along the fiber target could be studied offline. 
During deceleration the beam is lost
and the synchrotron prepared for the next cycle. The SEM
signal indicates that the luminosity roughly remains constant during
beam acceleration since beam losses are compensated by the
increased beam-current when the revolution-frequency raises from 0.5
to 1.5\MHz.  
{\em It should be noted that 
the beam emittance growth due to small angle scattering
in the fiber target
was nearly compensated by the adiabatic damping
due to beam acceleration \cite{hin89}.} During the flattop 
the luminosity decreases with a lifetime in the order of
5-10~s.

After retraction of the target, a laser monitor system (cf. next
section) is activated for 2\s in 
the flatbottom part of the cycle. For calibration and beam diagnostics, 
alternative cycle modes are applied, e.g. measurements at fixed momenta
with longer flattop for detector calibration, cf. \Sec{vertex}.

\subsection{Laser Monitor System}\mylabel{laser}
The EDDA-detector was monitored continuously by a laser monitor
system. Light pulses are generated at a rate of 20~Hz by a nitrogen
laser which drives a dye laser. This produces light pulses with
mean wavelength ($\lambda$ = 425~nm) and pulse shape characteristics 
similar to those generated in the scintillators by real particles.
They are fed through individual fibers into the light guides of all
scintillator elements to illuminate all photo-tubes simultaneously for
an online detector control. Pulse-to-pulse variations in laser
intensity are accounted for by normalization to photodiodes receiving
a constant  fraction of the dye laser light. In offline analysis the 
laser events are used to detect long term gain drifts or jumps between
detector calibration runs and to deduce necessary corrections.   

For very low energy ejectiles the produced light may exceed the linear
range of the PM-response. To this end, separate laser runs are
performed with the pulse intensity being stepwise decremented by a set
of calibrated optical filters. By comparison to the photodiode
reference the functional dependence of the digitized collected charge
from the PM on the light intensity is mapped out. Offline these
correction functions can be used to linearize the response of all
photo-tubes individually (\cite{lindlein96,COF97} and \Sec{vertex}).  

%
%
\subsection{Trigger and Data Acquisition (DAQ)}\mylabel{trigger}
%
The structure and granularity of the scintillator hodoscope reflect
the signature of \pp\ elastic scattering events, namely (i) coplanarity 
\begin{equation}
|\myphi_{2} - \myphi_{1}| = 180\grad
\mylabel{eq:1}
\end{equation}
and (ii) kinematic correlation of scattering and recoil angle, viz
\begin{equation}
\tan\Thlabi{1} \cdot \tan\Thlabi{2} = \frac{1}{\gcm^2},
\mylabel{eq:2}
\end{equation}
where $\gcm = \sqrt{1+\Tp/2m_pc^2}$ denotes the Lorentz
factor of the \pp\ center--of--mass motion as a function of beam
kinetic energy \Tp{}. Both conditions are used
to define a fast online trigger. The coplanarity, \Eq{eq:1},
is verified by requiring a coincidence of two scintillator bars
just opposite to each other with respect to the beam. Light
attenuation along the bars is taken care of by adding the readouts
from either end. The different flight and light transport times as
well as electronic constraints require a sufficiently wide (70\nsx)
coincidence gate. Monte Carlo simulations confirm that the width of the bars
guarantees a trigger even for deviations of the vertex position by  
as much as $\pm$1.5\cm from the nominal COSY beam axis.

The trigger condition for elasticity, \Eq{eq:2}, correlates
each semi--ring on one (e.g. left) side of the beam with one or more on the
opposite (e.g. right) side.
These correlations, however, depend on projectile energy
\Tp\ as well as the transverse profile of the beam-target overlap.
If the range of rings allowed in coincidence is sufficiently wide to
be valid for the whole momentum range covered 
during beam acceleration only a poor suppression of quasi-elastic
scattering in the carbon-nuclei is achieved. For this reason
programmable logic modules have been used to implement the
coincidence, such that they can be reprogrammed within 5\ms while the
beam is accelerated. A gain in trigger efficiency by a factor of two
(cf. \Fig{fig:trigger}) was achieved by reprogramming three times during acceleration, 
sacrificing 1\perc of data acquisition time. The times when the
reprogramming occurred was changed frequently to avoid empty spots in the
final excitation functions.

\begin{figure} 
\begin{center}
\includegraphics[width=\figwidth,bb=0 5 520 370]{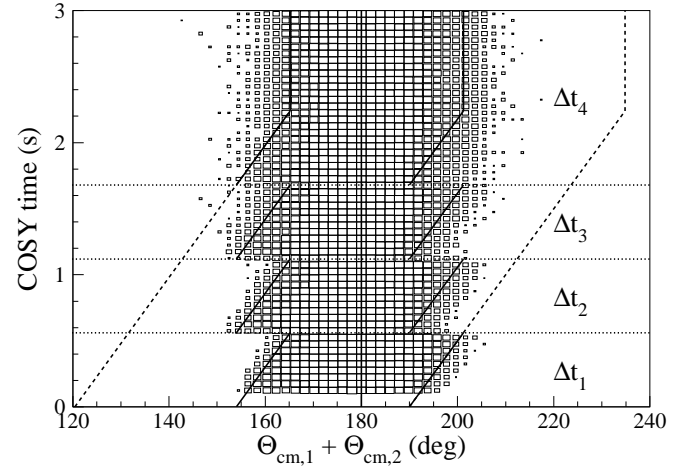}
\end{center}
\caption{Scatterplot of events taken with a carbon target during acceleration. 
         By threefold reprogramming (dotted horizontal lines) of the
         kinematic trigger, the reduced  
         acceptance (solid lines) suppresses uncorrelated background that 
         cannot be avoided with a fixed setting (dashed lines).}
\mylabel{fig:trigger}
\end{figure}

Data are acquired whenever the target is in the beam or the
laser system is triggered. Data are processed with conventional CAMAC
modules for timing, pulse-height and logic signals in conjunction with
a VME based event-builder. An event is written on tape when \Eqs{eq:1}{eq:2}
are fulfilled by at least one pair of scintillator bars and
semi--rings, respectively.  
In addition, each event contains the cycle time from a COSY clock controlling 
the accelerator operation, the collected charge from all scintillator
elements and the timing signals from both ends of the scintillator bars.

There are three additional trigger sources, namely for PIN and SEM
luminosity monitors and for the laser system, which are merged with
the main trigger. All four independent trigger sources are counted,
together with a 10\MHz clock and pulses counting the instantaneous
frequency \fhf\ of the COSY cavities, in scalers which are read
out in time increments of 2.5\msx. Comparing the number of stored
events with the scaler of the corresponding trigger source yields the
dead time fraction $\tau(\mom)$ of the data acquisition system as a function of
beam momentum (i.e.\ cycle time) 
that can be corrected for in beam momentum intervals as narrow as
3\MeVc\ (cf. \Sec{deadtime}). Our measurements were performed with
time--averaged luminosities of typically $5\cdot 10^{29}\cmx^{-2}\sx^{-1}$ and
singles rates of up to 40\kHzx.  
%

%
\section{Offline analysis}\mylabel{offline}
%
\subsection{Data Samples}\mylabel{samples}
%
The data were collected in two production runs of two and three weeks,
separated by half a year, with a yield of about 
15$\cdot$10$^6$ and 22$\cdot$10$^6$ elastic events,
respectively. After publication of the results based on the first
production run \cite{albers97},
small modifications concerning e.g.\ the treatment of artefacts due to
the detector granularity have been implemented in the analysis,
advanced methods to reduce non-elastic background were developed, and
a correction for radiation damage of the \CH-target introduced.
Furthermore, the multitude of different running conditions,
have allowed additional consistency checks and
both data sets have been (re--)analyzed for the present work.

First, the data have been divided in 17 so-called data samples, which
have been analyzed separately. 
Each sample comprises typically $1.5\ldots4\cdot10^{6}$ elastic events and
consists of two data sets, one taken with a \CH-- and a C--target each. The
relative luminosity of the \CH-- and C--target data, measured
interleaved by changing targets every few hours, was chosen online 
to yield about the same statistics for proton-carbon scattering events.
Each data sample contains data 
\begin{enumerate}
\item from only one \CH - and one C-target,
\item with exactly the same trigger setting (e.g. reprogramming times)
\item and the same detector and COSY-beam setup.
\end{enumerate}
By treating the individual results for these
samples as separate measurements many consistency checks could be
made prior to combining them to the final results (see \Sec{consistency}).

\subsection{Overview}\mylabel{anaov}
%
For $N_{pp}$ detected elastic scattering events within a solid angle
bin $\Delta\Thcms$ and a momentum bin $\Delta \mom$, centered at
(\mom, \Thcmsx), the differential 
cross section is given by
\begin{equation}
\frac{d\sigma}{d\Omega}(\mom, \Thcmsx) = \frac{N_{pp}(\mom,
\Thcmsx)} 
{{\Delta\Omega} \cdot \eta(\mom,\Thcmsx) \cdot L_H(\mom) \cdot {(1-\tau(\mom))}
}. 
\mylabel{eq:2A}
\end{equation}
Here, $L_H$  denotes the absolute
luminosity and $\tau$ the DAQ dead time fraction for the respective
momentum bin.   
The detection efficiency $\eta$ accounts for the elastic scattering
events discarded by the trigger or offline cuts, mainly due to
secondary reactions of scattered protons in the experimental setup
(\Sec{efficiency}). 

For elastic proton-proton scattering we detect two protons and a rough
selection of events is based on detector multipli\-cities. To this end,
we subdivide the data in event classes based on the
number of hits in the scintillator elements of the
hodoscope and keep only those of low multiplicity. 
The scattering angle \Thcms is reconstructed from the
vertex position and the interception points of the two trajectories,
determined from the energies deposited in the scintillator
elements. The vertex position moves along the target fiber during beam
acceleration. Applying a kinematic fit, it can eventually be
determined event\-wise.  
The corresponding concepts are described in \Sec{vertex} and
were applied to  
measurements with both types of target fibers.

For background reduction, a combination of an event-wise selection
based on \Eqs{eq:1}{eq:2} and a statistical
subtraction of the carbon fiber data, normalized to the same
luminosity, is accomplished in a way that -- in first order --
inelastic \pp\ interactions are corrected for as well. Details are
given in \Sec{cuts}. For all 
excitation functions, the {\em relative} luminosity is provided by the
two luminosity monitors. The determination of the
{\em absolute} luminosity $L(\mom)$ in this experiment is not as
accurate as in some high-precision external experiments, so that
we normalize all excitation functions at one single momentum \mom\ to published
data (cf.\ \Sec{lumi}).

\subsection{Monte Carlo simulations}\mylabel{mc}

Experiment and analysis are accompanied by a comprehensive
Monte--Carlo simulation including the details of detector geometry and
materials. Particles are tracked through the detector, taking into
account energy and angular straggling as well as hadronic and
electromagnetic secondary interactions. Results of Monte-Carlo studies
enter mainly threefold into the analysis:
\begin{enumerate}
\item Calculating the efficiency $\eta$ for accepting elastic
\pp-scatter\-ing events, this includes effects of the trigger as well of
all software cuts used in the analysis.
\item For modeling the contribution of inelastic \pp-reactions in order 
to estimate the background contribution to accepted \pp-elastic events.
\item To calculate the detection efficiency of the PIN-diodes for
$\delta$-electrons used for measuring the relative luminosity.
\end{enumerate}
For electromagnetic interactions we use EGS4 \cite{EGS4} and
for hadronic reactions MICRES \cite{Ackerstaff:2002hj}
or phase-space distributions as event generators. 
In MICRES a number of  event generators are used: To model
elastic scattering cross sections distributions according to
recent phase-shift analysis \cite{Arndt:1997if} are
generated. Inelastic scattering 
of hadrons on nucleons and 
nuclei for energies up to 5\GeV\ are described by tracking the
impinging nucleon through the nucleus. Secondary particles
including pions and hadrons are traced and their interactions with
nucleons are described with known, interpolated or estimated cross
sections. Momentum, energy and charge are fully conserved.
In this intranuclear cascade the
nucleon--nuc\-leon interactions are either quasi-elastic or
inelastic, which is modeled by exciting resonances, which may
subsequently decay and eventually produce mesons.  
All nuclear resonances contributing more than 2\perc to the total
inelastic cross section are taken into account. 
 Details of this event generator MICRES are given elsewhere
\cite{Ackerstaff:2002hj}. 
%
%
%
\subsection{Multiplicity Cuts}\mylabel{multi}
%
For a first classification of an event, a group of $n$ hits in
adjacent scintillator bars or semi--rings is called a cluster of size
$n$. Due to the geometrical overlaps (cf. \Fig{fig:elements}) cluster
sizes $n$ = 2 and 3 dominate. An event can then be characterized by
its cluster pattern (N$_B$, N$_L$, N$_R$), where N$_B$ is the
multiplicity (i.e.\ the number) of clusters in the layer of
scintillator bars, and N$_L$ (N$_R$) denotes the respective
multiplicity for the left (right) group of semi--rings. An ideal
elastic \pp\ event reveals the pattern (211). The majority of elastic
scattering events is of this type. There are, however, several
patterns of higher cluster multiplicity that originate from elastic
scattering due to e.g.\ secondary reactions in the scintillator
hodoscope, accompanying $\delta$--electrons or random
coincidences. By comparing the \CH and C-data sets the patterns
containing non-negligible numbers of elastic events could be identified 
and we kept eight patterns, namely
(211), (311), (411), (221), (212), (321), (322) and (222). For a
projectile energy \Tp\ = 1.5~GeV, these patterns exhaust 98.4\perc
of all triggered events, with the ideal (211) pattern already
contributing 89.2\percx. 
All other patterns are discarded.

%
%
\subsection{Trajectory and vertex reconstruction}\mylabel{vertex}
%
EDDA primarily detects the position of the protons on cylinders 
surrounding the beam pipe. The coordinate along the circumference
$R\cdot\myphi$ is determined from the scintillator bars (with $R = 16.4$ cm) and
the $z$ position along the beam by the semi-rings. 
The resolution can be enhanced considerably beyond that given by the
detector granularity by evaluating the charge-inte\-grated
photomultiplier-signal $S$ of all scintillator elements, usually two,
hit by the particle.
This can be related to the pathlength within the scintillator element $\Delta
x$ and to the average specific energy loss along this pathlength
$\overline{(\dE/\dx)}$ by
\begin{equation}
S(\Thlabx, \myphi) = G \cdot \overline{\left(\dEdx\right)} \cdot\Delta x 
\mylabel{eq:3}
\end{equation}
For most projectile energies and scattering angles $\Thlab$ covered in the
experiment, the energy loss can be considered constant along the whole path 
through adjacent, overlapping scintillator elements. In this simple
case the gain factor $G$ describes the conversion of deposited energy
to the electronic signal and encompasses  
(i) light transportation and attenuation in the
scintillator and light guide, and (ii) photomultiplier (PM) response.
The former effect introduces a dependence on the particle's point of
incidence $G(\Thlabx, \myphi)$ which is mapped out experimentally by 
\pp\ elastic scattering data taken at fixed momentum \mom~=~2.7~\GeVc, where
the detector acceptance is at its maximum.  

For the latter, a
laser-light (cf. \Sec{laser}) fed into the light guides was used to
deduce corrections for long-term  gain drifts and nonlinearities close
to the upper limit 
of the PM's dynamic range.

The spatial resolution in $\myphi$ (\Thlabx) can then be enhanced beyond the 
granularity given by the number of scintillator bars (semi--rings),
by interpolating the point of incidence using the pulse heights $S_1$,
$S_2$ of neighboring scintillators traversed by the charged
particle. The ratio  
\begin{equation}
Q_{12} = \frac{S_{2}/G_{2} - S_{1}/G_{1}}{S_{2}/G_{2} + S_{1}/G_{1}} 
\mylabel{eq:4}
\end{equation}
varies according to \Eq{eq:3} monotonically with 
$\Delta x_{2} - \Delta x_{1}$ between +1 and $-1$ along the interval
of overlap, see \Fig{fig:elements}. Details of the algorithm are given in
\cite{bisplinghoff93}. The resolution obtained for an ejectile is improved by a
factor $\approx 5$ to $\delta\myphi \approx 1.9\grad$ (FWHM) and
$\delta\Thcms \approx 1.0\grad$ (FWHM). 

\begin{figure}
\begin{center}
\includegraphics[width=\figwidth,bb=0 0 520 800]{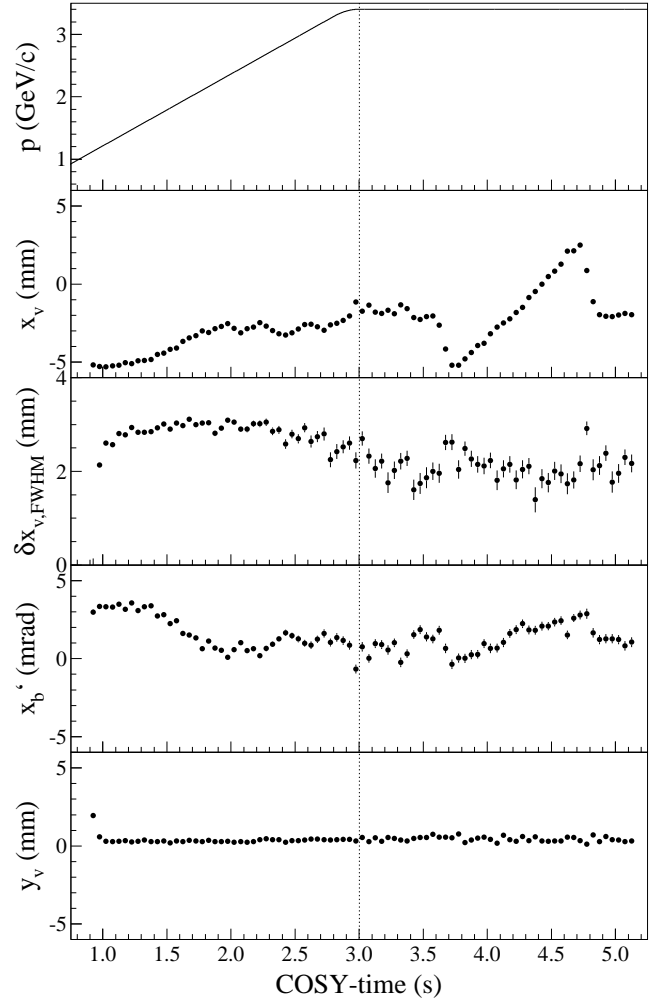}
\end{center}
\caption{Variation of horizontal vertex parameters and vertical beam
position $y_v$ as a function of cycle time. The vertical line shows
the begin of the flattop where the beam is swept on purpose horizontally
over the fiber target. 
$x_v$ is the beam position, $\delta x_v$ the beam width folded with the
detector resolution, $x'_b$ the horizontal beam angle and $y_v$ the vertical
target position. The first two points in $y_v$ show the settling of
the target after it has been moved in.
}
\mylabel{fig:vertex}
\end{figure}

For ejectile energies below 400~MeV two corrections are important:
First, the specific energy loss \dE/\dx\ can no longer be considered
constant along the path of a particle. 
Assuming that the particle is an elastically scattered proton, so that
the kinetic energy is known and
taking into account the stopping power of all material in
front of the detector element under consideration a correction can be
deduced. Secondly, the amount of produced light is affected by
nonlinearities in PM response. Therefore, it is linearized using
information obtained with the laser monitor system (cf. \Sec{laser}).
With these corrections a good angular resolution is maintained 
except for the lowest energies, where systematic deviations of the
reconstructed position arises which will be treated as a systematic error.

In order to obtain the scattering angle the reaction vertex must be
known. The vertical and longitudinal position are fixed by the fiber
position and are thus the same (within a few $\mu$m) for all events.
The horizontal vertex position is smeared out by the width of the COSY
beam and should be determined for each event separately.
As a first step the first moments of the vertex distribution, e.g. the
mean values $(x_v,y_v,z_v)$ and the horizontal width $\delta x_v$ are
deduced from the data, by exploiting the coplanarity of elastically
scattered protons with the beam or the kinematic correlation.

The vertex position $x_v$, $y_v$ in the plane perpendicular to the
symmetry axis of the detector can be determined from the pairs of
points where the prongs intercept the scintillator bars. Due to the
coplanarity of two body interactions the projection of the line
connecting the two points into this plane must cross the vertex. The
elastic scattering events are evenly distributed in $\myphi$; the
coordinates $x_v$, $y_v$ are obtained as parameters from a fit to all
events for a given projectile momentum interval $\Delta \mom$ (details
are given in \cite{Rohdjess:2004b}). The
method can be extended to projectile beams tilted against the detector
axis. By comparing events with equal ($\Thcmsx \approx
90\gradx$) or different ($\Thcmsx \ll 90\gradx$) lab scattering angles
yields the directions $x'_{b}$ and $y'_{b}$ as well
(cf. \Fig{fig:vertex}).  
The validity of this method was tested and verified with a beam of
fixed momentum that was steered in a  
controlled way horizontally ($x$) along the target fiber ($3.6$~s~$< t
< 4.8$~s in \Fig{fig:vertex}).
The horizontal vertex position and the tilting angles of the beam
turned out to vary during acceleration to a non negligible
extent. These parameters, however, showed a remarkable 
long term stability over many days.
Finally, the vertex position $z_v$ in beam direction is calculated from the interception 
coordinates $z_1$, $z_2$ in the ring-layer by making use of the
kinematic relation 
\Eq{eq:2} with $\gcm$ being known from the instantaneous 
projectile momentum and $\tan\Thlabi{i} = R/z_i$, where $R$ is the
mean radius of the semi-ring layer.  
This way, the vertex coordinates $x_v$, $y_v$, $z_v$ are determined with an 
accuracy $\pm$ 0.5\mmx; the width $\delta {x_v} \approx 3\mmx$ (FWHM) 
reflects the horizontal beam width folded with the experimental resolution. 

With the mean vertex position and the position of the hit in the detector known,
the polar and azimuthal scattering angles are calculated, taking into
account the small correction due to the tilted beam.
With the reconstruction algorithm described above, however, the
horizontal vertex position cannot be determined {\em
event\-wise}. To improve the resolution of the final \cms\ polar
scattering angle \Thcmsx, it is taken
from a kinematic fit, constraint by
elastic scattering kinematics (i.e 
\Eqs{eq:1}{eq:2}). In the fit $x_v$ is treated as a free parameter; its
distribution is compatible with the previous result for $\delta x_v$.

For events with higher multiplicities, i.e. pattern other
than 211, the ambiguities when correlating hits from the bar and the ring
layer can be resolved almost entirely by the time-difference of the signal read
out at the down- and upstream sides of the scintillator bars, which
yields a position information along the bar of about 5~cm (FWHM) and
must match the position of the struck rings.
If ambiguities remain, the combination of hits which best matches 
elastic scattering kinematics, as outlined in the next section, is selected.

\subsection{Event Selection and Background Subtraction}\mylabel{cuts}
%
\begin{figure}
\begin{center}
\includegraphics[width=\figwidth]{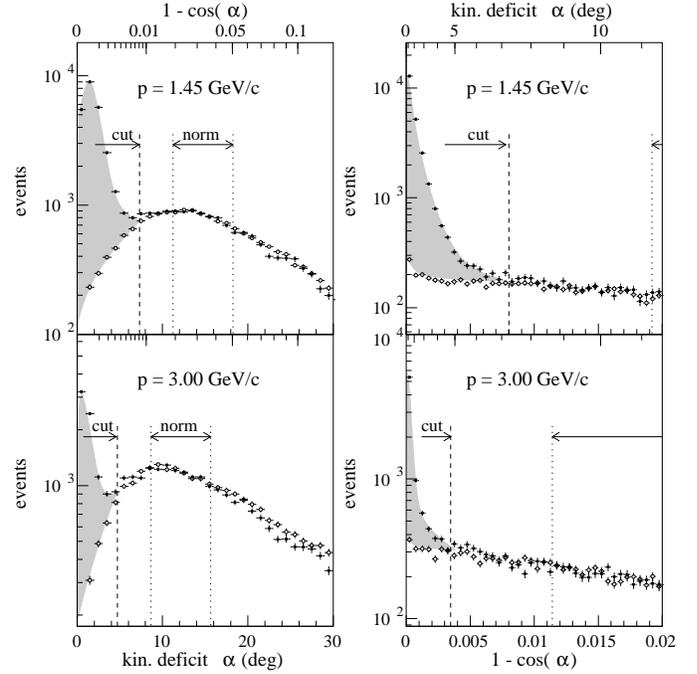}
\end{center}
\caption{Distribution of the kinematic deficit $\kd$ for two beam
momenta. Shown are events from  \CH (closed) and C-targets
(open symbols). The elastic \pp\ events are indicated by the shaded
area. The upper limit $\kdc$ (cut) for accepted
elastic scattering events and the range (norm) used for normalizing the
C-target data to the \CH -data are shown. 
The solid angle associated with \kd\ rises with
$\cos\kd$, so that the number of events at very small \kd\ drops. The
figures on the right show that indeed the elastic peak is at $\cos\kd
= 1$ as expected. Note the logarithmic scale.
}
\mylabel{fig:alpha}
\end{figure}
%
For each event we have at least two ($i = 1,2$) reconstructed tracks with 
angles (\Thlabi{i},\Ph{i}) that are now subject to additional cuts
to separate elastic \pp\ scattering events from inelastic reactions or
background originating from proton-carbon reactions.
For this purpose one first calculates the \cms\ scattering angle using
\pp\ elastic scattering kinematics:
\begin{equation}
\Thcmsi{i} = 2\,\arctan\left(\gcm \tan\Thlabi{i}\right);
\mylabel{eq:thcmdef}
\end{equation}
Note that $\myphi = \myphi_{\rm lab} = \myphi_{\rm \cms}$.
For elastic events both particles should be emitted back-to-back in the
center-of-mass system. The angular deviation $\kd$ from this perfect
180\grad correlation, subsequently called the {\em kinematic deficit}, 
is shown in \Fig{fig:alpha} for \CH-- and C--target data at two
different beam momenta. The kinematic deficit is used twofold:
First, for the few events with more than two prongs the combinatorial
ambiguity is removed by selecting the combination with minimum
$\kd$. 
The resulting probability distributions $N(\mom,\kd$) show that
elastic \pp\ events clearly stand out  
on a smooth, monotonic background. 
A comparison to the distribution obtained with a pure 
carbon fiber target under identical experimental conditions shows
that this background can essentially be associated with
quasi-free scattering on bound protons and inelastic p-C
interactions. 
A display of events more closely
related to the conditions of coplanarity, \Eq{eq:1}, and
elasticity, \Eq{eq:2}, is shown in \Fig{fig:thetaphi} where the
elastic peak extends about two orders of magnitude above the
background. However, the combination of those two requirements in one
variable, $\kd$, is more effective for quantitative background
treatment. 

\begin{figure}
\begin{center}
\includegraphics[width=\figwidth]{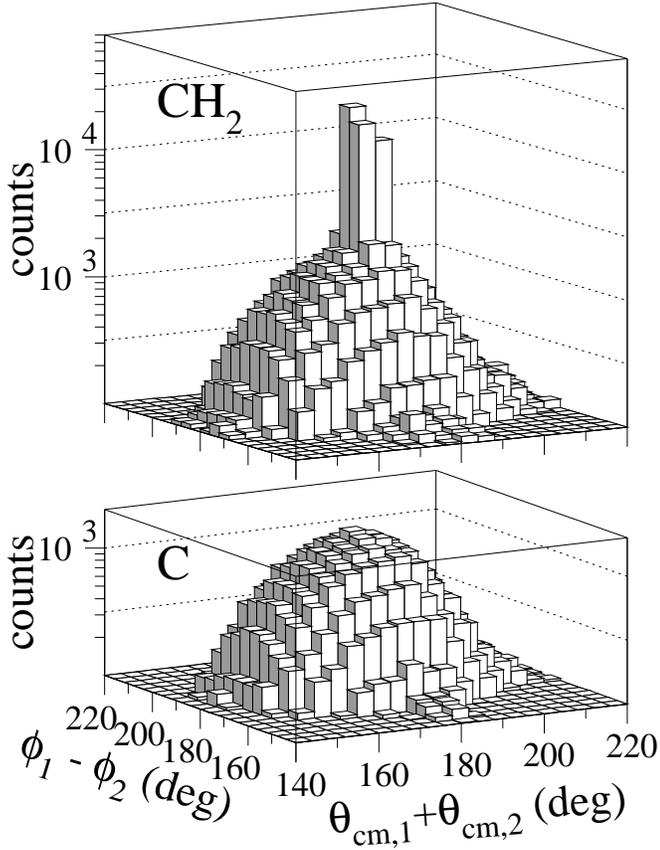}
\end{center}
\caption{ Distribution of angle integrated events obtained with a
\CH\ (top) and carbon (bottom) fiber target at 
\mom~=~2.25~\GeVc. Note the logarithmic scale.}
\mylabelnr{fig:thetaphi}
\end{figure}

In a first step, a projectile momentum dependent cut $\kd \leq
\kdc(\mom)$ is applied in a way that -- based on Monte-Carlo
simulations and comparisons to the C-target data -- almost no elastic events
are discarded. The remaining events 
are sorted into two--dimensional arrays
\Nch(\mom,\Thcmsx) and \Nc(\mom,\Thcmsx) with
the option to adjust the bin-widths $\Delta \mom$ and $\Delta
\Thcms$ to statistical requirements. 

\subsubsection{Statistical Background Subtraction}\mylabel{carbonsubtract}
Background subtraction is performed statistically: A factor \fnorm(\mom)
is deduced from the normalization in the interval $\kdmin = \kdc +
4\grad\ \leq \kd \leq \kdc + 11\grad\ = \kdmax$ of the
$\kd$-distribution  \Nch(\mom,\kd) of a \CH--target data set to
that \Nc(\mom,\kd) of the associated C--target data set, i.e.
\begin{equation}
\fnorm(\mom) = \frac{
\left.\Nch(\mom) \right|_{\kdmin\leq\kd\leq\kdmax}
}{
\left.\Nc(\mom) \right|_{\kdmin\leq\kd\leq\kdmax}
}
\mylabel{eq:csublum}
\end{equation}
where events have been integrated over all scattering angles.
\fnorm\ can be viewed as the luminosity with respect to proton-carbon
scattering of the \CH--target relative to the C--target data set. Its momentum dependence is
caused by the different COSY beam lifetimes for the two targets due to
their different thickness. \fnorm\ was determined for all contributing
multiplicity 
patterns separately, to account for slightly modified contributions from 
accidental coincidences caused by the luminosity difference.
The background can now be subtracted statistically viz
\begin{equation}
\Npp{pp}{}(\mom,\Thcmsx) = \Nch(\mom,\Thcmsx) - \fnorm(\mom) \cdot
\Nc(\mom,\Thcmsx) 
\mylabel{eq:csub}
\end{equation}
There are two concerns when applying this method: first, the
\CH--targets are aluminum-coated and it must be verified that p-C and
p-Al reactions in this region of phase-space are indistinguishable.
This was proven by taking data sets with plain and aluminum
coated C--targets. The carbon-subtracted \pp\ elastic scattering yields
is the same within statistical errors. Applying the statistical
subtraction scheme to C+Al-- and C--target data sets (instead of \CH\
and C) produced results for ``\Npp{pp}{}{}'' compatible with zero.
A second concern is that inelastic proton-proton scattering populates 
the \kd-distribution in the region where \fnorm\ is determined and
will therefore systematically distort the results. To shed light on
this issue Monte-Carlo simulation were performed to mimic the effect
of inelastic \pp\ reactions.

\subsubsection{Contributions of Inelastic pp-Reactions}\mylabel{inela}
Elastic scattering exhausts at least half the \pp\ total cross section
in the COSY energy range. Except for pp$\rightarrow$d$\pi^+$ all
inelastic reactions have at least 3 particles in the final state and 
therefore the probability to look like an elastic
event is expected to be small. 
Since these reactions cannot be measured separately by our experiment
we rely on Monte-Carlo (MC) simulations, using the available 
information on the respective total cross sections. 
Unfortunately, phase-space distributions are practically unknown so
that models have to be used in MC event generation. For this purpose
distributions either according to phase-space or generated by MICRES,
where inelastic reactions proceed through excitation of resonances, are used.

Simulations show that the main inelastic contribution originates from
the channels 
pp$\rightarrow$ pn$\pi^{+}$, $\rightarrow$ pp$\pi^{+}\pi^{-}$ and
$\rightarrow$pp$\pi^{\circ}$.  
The cuts on the cluster pattern and the correlation mismatch $\kd$ reduce 
this background to less than 1.5\% of the (angle integrated) elastic yield, 
whereas the corresponding number for the p-C background is up to one order 
of magnitude higher; both increase with beam momentum \mom. 

\begin{figure}
\begin{center}
\includegraphics[width=\figwidth,bb=2 2 730 280]{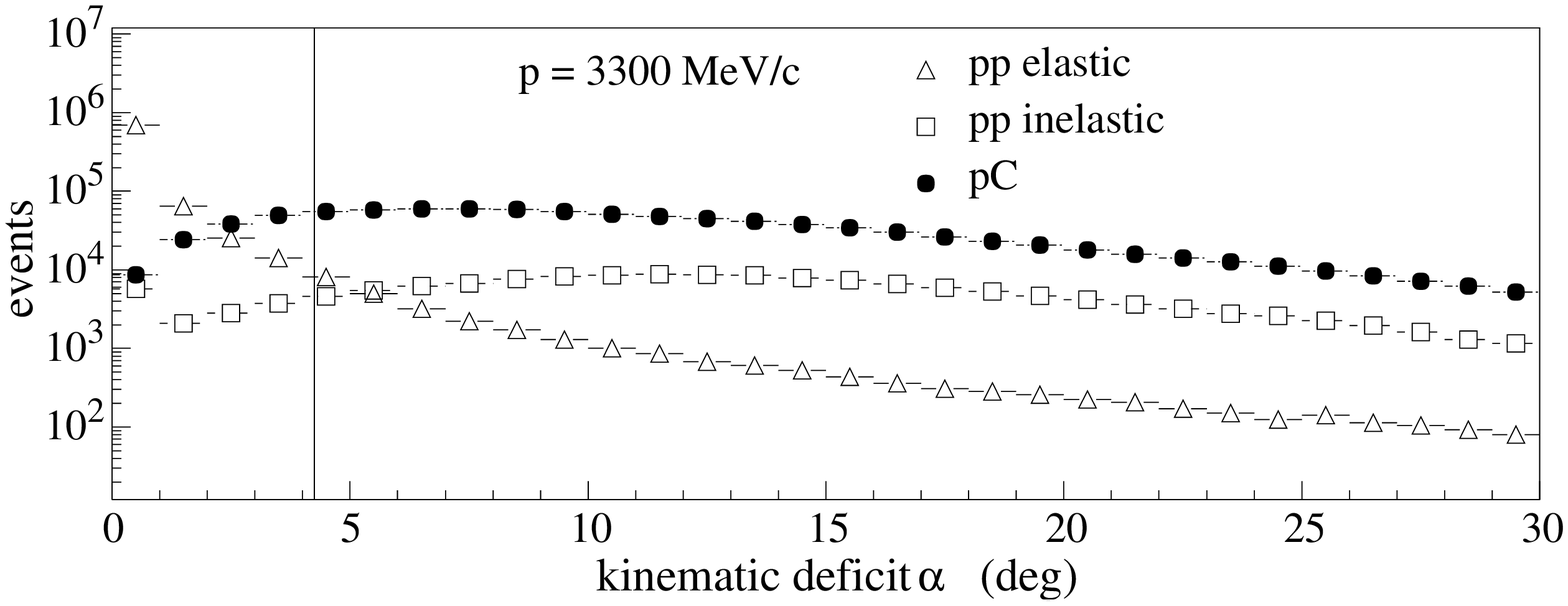}
\end{center}
\caption{ Simulated distributions of the kinematic deficit $\kd$ for 
          p-C reactions (solid dots), elastic (triangles) and
inelastic (squares) \pp\ scattering. The vertical line indicates the
position of $\kdc$.} 
\mylabel{fig:alphaSim}
\end{figure}

Their simulated $\kd$-distributions are shown in \Fig{fig:alphaSim}
for the worst case, i.e. at high momenta. Both background
distributions 
are similar in shape. Therefore, the statistical subtraction with
its normalization performed in the angular range $\kd \geq
\kdc$ takes care of the main part of \pp\ inelastic events
with $\kd \leq \kdc$, too. A closer look, however, reveals
differences in shape that will give rise to a systematic background
error which has been studied in more detail.

For this purpose, $\Nch(\mom,\Thcmsx)$ has been numerically
composed from three contributions, namely 
\begin{equation}
\NCH = \Nppel + \Nppin + \NCa{a} 
\mylabel{eq:inel}
\end{equation}
where the elastic contribution \Nppel\ is based on our
experimental results \cite{albers97}. The second term is Monte Carlo
generated and the third is an experimental data sample \NCa{a}\ obtained with
a carbon target, properly normalized to the observed ratio of \pp\
to p-C events. This set is then corrected for background with the
statistical method, \Eq{eq:csub}, by using a second,
statistically independent
carbon measurement \NCa{b}. From the result the term
\Nppel\ used before in \Eq{eq:inel} is subtracted. The
remaining difference $D(\mom,\Thcmsx)$ reflects the incorrect
treatment of the inelastic \pp\ component \Nppin, and turns out to be
systematically {\it positive}, indicating that the inelastic \pp\
contribution was 
insufficiently subtracted as is expected from \Fig{fig:alphaSim}. 
The fraction $\left|D/\Nppel\right|$ is negligible (i.e. less than 1\%)
except for the large-angle ($\Thcmsx > 70\gradx$) and high-momentum
($\mom > 2.5~\GeVc$) region where the
contribution may reach about 5.5\%. Because only the total
cross sections but not the angular distributions of most inelastic
\pp\ reactions are sufficiently well known, no attempts have been made 
to correct this deviation and the maximum effect obtained with
different models is taken as a systematic uncertainty.

\subsubsection{Hydrogen Absorbed in Carbon-Fibers}   
A close look at the distribution for events measured with carbon-fiber
targets reveals a tiny peak at the locus of elastic \pp\ scattering,
cf. \Fig{fig:thetaphi}, which 
could be due to traces of hydrogen or water absorbed in the carbon
fiber. Indeed does this peak 
show an angular distribution in agreement with the expectation for
elastic \pp\ scattering. The background subtraction following
\Eq{eq:csub} will then reduce the number of elastic \pp\ events of the 
p-\CH\ measurement systematically wrong. A careful study of this
reduction showed that it is for all targets smaller than 0.5\% and
does not vary with projectile momentum. Therefore, it is
equivalent to a reduction of the effective \CH\ target thickness and
its effect cancels completely when finally all excitation functions
are normalized at one momentum to some precise reference cross-section, cf.\
\Sec{lumiAbs}.

\subsubsection{Statistical Errors}\mylabel{carbonSubError}

\begin{figure}
\begin{center}
\includegraphics[width=\figwidth]{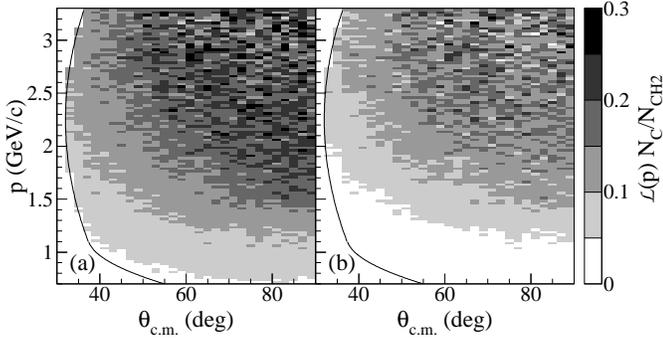}
\end{center}
\caption{(a) Ratio of events from carbon- and \CH-targets of one data
sample after all cuts have been applied and the carbon data have been
properly normalized. The solid line shows the
acceptance cut on the 
c.m. scattering angle. In (b) the same data are shown when in addition
event selection using fuzzy-logic methods is utilized.
}
\mylabel{fig:cContribution}
\end{figure}

The {\em statistical} error associated with the background subtraction
is roughly given by $\delta\Npp{pp}{} = \sqrt{\Npp{\CH}{} + \fnorm^2
\Npp{C}{}}$ where the error of \fnorm, which is a minor contribution,
has been left out for the sake of simplicity. Thus any contribution 
from scattering off carbon that can be eliminated by cuts prior the
subtraction procedure (\Eq{eq:csub}) will reduce $\delta\Npp{pp}{}$. In
\Fig{fig:cContribution}~(a) the fraction of events in $\Npp{\CH}{}$
attributed to the carbon-content of the target is shown.  
It rises with momentum and angle from 5\% up to 25\% and unfortunately
is largest where scattering rates are the lowest. 
Therefore, it was investigated how the statistical method
could be complemented by an event\-wise classification scheme to reduce
\Npp{C}{} prior to being subtracted. To this end information on each
event besides the measured angles, like energy losses, can be
used. When testing methods commonly employed in pattern-recognition,
like self-organizing maps or artificial neural networks, we found
an approach based on so-called fuzzy-logic to be particularly
helpful. It allowed a large reduction of 
the number of proton-carbon scattering events to be subtracted
statistically as shown in \Fig{fig:cContribution}~(b), and will be
discussed in detail in the next section.

\subsubsection{Background subtraction using Fuzzy-Logic}
So far only the information on the proton angles have been used to
identify elastic scattering events. In addition, the EDDA detector
provides energy-loss information in the bars and semi-rings and timing
information from the bar layer, which can be put into use to test
their compatibility with elastic \pp-scattering events. However,
quasi-elastic scattering from nucleons in the carbon-nucleus as well
as inelastic \pp\ reactions are not cleanly separated in these
variables, so that simple cuts are insufficient. Here, methods
developed
within the framework of so-called ``fuzzy-logic''
\cite{bothe95,hoppner1999} have been shown to
be very powerful  \cite{busch01}.

For each event, described by a vector \vecQ\ of $n$ measured
or reconstructed quantities $q_i$ ($i = 1\ldots n$), we must construct
a decision-function $d(\vecQ)$ to decide if it is accepted
($d(\vecQ)=1)$ or not $(d(\vecQ) = 0)$  as a candidate for an elastic
scattering event. When using simple cuts on one or two-dimensional
projections of $\vecQ$ applied in sequence most of the correlations
between the observables $q_i$ are lost, since one variable outside one
cut is sufficient to discard an event. In the fuzzy-logic approach we
define for each observable $q_i$ a so-called member-function 
$P_i: q_i\mapsto [0,1]$, which maps the parameter space of $q_i$ onto the
interval between zero and one. It is designed such, that events with
a high (low) probability of being elastic yield  values $p_i$
close to 1 (0).  
For this purpose the member function must not be a probability in the
strict sense and we use
\begin{equation}
P_i(q_i) = 1-\frac{\NC(q_i)}{\NCH(q_i)}.
\mylabel{eq:fuzzy1}
\end{equation}
To obtain $P_i$ with sufficient statistical precision, the data are
binned in $q_i$. \NC\ and \NCH\ are the number of events within a
certain bin centered at $q_i$ for this quantity, whereas all other
quantities $q_j$ with $i \neq j$ may have arbitrary values.
Note, that here and in the following it will be assumed that the \CH--
and C--target data samples have been properly normalized to the same
luminosity.

\begin{figure*}
\begin{center}
\includegraphics[width=\widefigwidth]{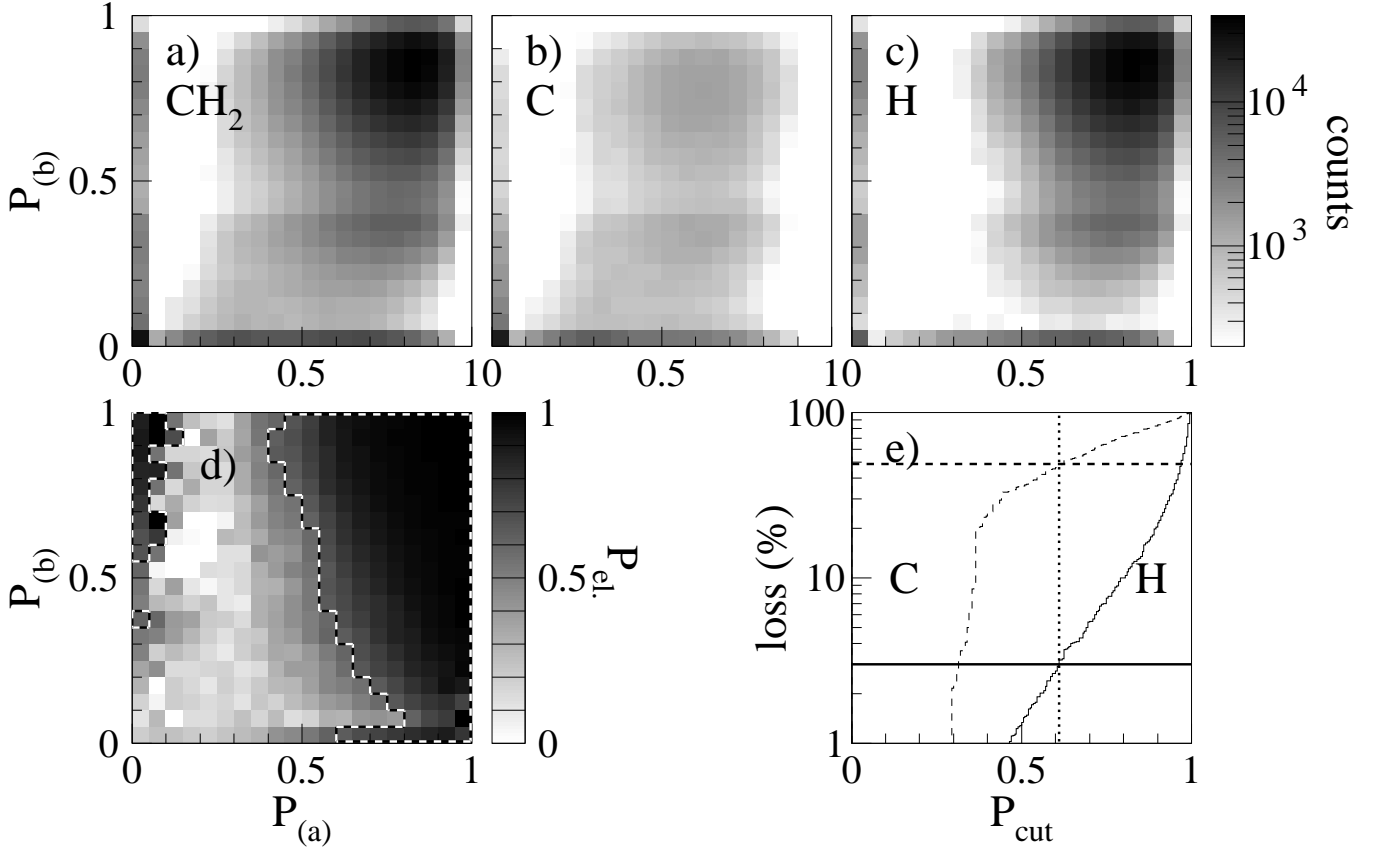}
\end{center}
\caption{Distribution of events in the (\Probpa,\Probpb) plane 
for a \CH~(a) and C-target (b) normalized to the same luminosity.
In (c) the difference attributed to \pp-scattering is shown, and in (d)
the ``probability'' \Pel\ as defined in \protect\Eq{eq:fuzzy3}. The
dashed line shows a typical contour for accepted events based on the
condition $\Pel~>~\Pcut$ which corresponds to a loss of 3\% of elastic
events (e) while reducing the background by about 50~\%. 
The spectra shown correspond to events around $p~\approx~2.1$~\GeVc\
and are integrated over all scattering angles.
}
\mylabel{fig:fuzzy1}
\end{figure*}

These member-functions are then combined to a combined ``probability'' 
by the $\gamma$-operator \cite{bothe95,hoppner1999} 
\begin{equation}
\Prob{\vecQ} = \left[\prod_{i=1}^n P_i(q_i)\right]^{1-g}\cdot 
\left[ 1- \prod_{i=1}^n\left(1-P_i(q_i)\right) \right]^g
\mylabel{eq:fuzzy2}
\end{equation}
mapping the parameters space in $\vecQ$ onto the interval $[0,1]$.
It features a, yet to be fixed, parameter $g\in[0,1]$, to the effect that the
$\gamma$-operator specializes to the well-known logical AND ($g =
0$) or OR ($g = 1$) operators in the limiting cases. 
By this procedure we have mapped the information contained in \vecQ\ 
to a single variable \Probp\ between 0 and 1.

In this experiment we used n=5 experimental quantities $q_i$:
\begindescription{ppppppp}
\item[$\kd$:] The kinematic deficit as defined in \Sec{cuts},
\item[\Dphi:] The difference of the azimuthal angles of the two
tracks,
\item[\DTZ] The correlation of the time-of-flight (TOF) $T_1 -T_2$
and the $z$-position difference in the bar layer of the two charged
tracks:
\begin{equation}\mylabel{eq:tofdiff}
\DTZ \equiv (T_1 - T_2) - \bcm\,\left(z_1 - z_2\right),
\end{equation}
where $\bcm =\sqrt{\Tp/(2m_p + \Tp)}$. This quantity is close to
zero for elastically scattered protons as required by kinematics.
\item[\DEez:]The energy loss in the bar layer for the particle
scattered to the left (right) of the beam.
\end{description}
These were combined to two vectors \vecQa = $(\kd,\Dphi,\DTZ)$ and
\vecQb = (\DEe,\DEz). This allows us to map each event 
onto a unit square spanned by \Proba{\vecQa} and \Probb{\vecQb}
as shown in \Fig{fig:fuzzy1} for $g=0.5$. The elastic events clearly
stand out in the upper-right corner in the (\Probpa,\Probpb) plane,
whereas reactions on carbon -- considered as background -- are
spread out over a larger region. 
Increasing the value of $g$ spreads out the background over a
larger area, whereas a decrease does the same to the elastic
scattering events, so that $g=0.5$ appeared to be a good compromise.

\begin{figure}
\begin{center}
\includegraphics[width=\figwidth]{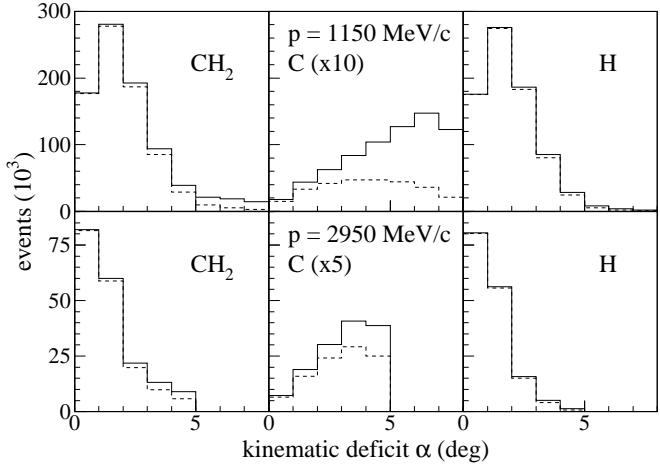}
\end{center}
\caption{Spectra of the kinematic deficit $\kd$ for two beam
momenta ranges ($\pm$150 MeV/c) with (dashed line) and without (solid
line) application of background rejection using fuzzy-logic. Shown are 
the data from \CH-- (left) and C--targets (center, normalized to the
same luminosity) as well as the difference (right) attributed to the
hydrogen contents. Note that a cut at $7.5\ldots 8\grad$ (top) or 
$4.5\ldots 5$\grad\ (bottom) on $\kd$ has already been applied.
}
\mylabel{fig:fuzzyAlpha}
\end{figure}

In principle one could now define the decision function
$d(\vecQa,\vecQb)$ by drawing a two-dimensional 
contour on the \Probpa\ vs. \Probpb\
distribution. However, the signal (\pp\ elastic scattering) and the
background (p-C reactions) do overlap, so that any reduction of the
latter will always be accompanied by some unavoidable loss of signal.
However, in large regions of the (\Probpa,\Probpb)-plane
the losses will be very small. This can be quantified by defining 
\Pel, again viz
\begin{equation}
\Pel(\Probpa,\Probpb) \equiv 1 - \frac{\NC(\Probpa,\Probpb)}{\NCH(\Probpa,\Probpb)}.
\mylabel{eq:fuzzy3}
\end{equation}
If we sort the data in 20$\times$20 bins in the (\Probpa,\Probpb)
plane (cf. \Fig{fig:fuzzy1}~(d)) we can view \Pel\ as a probability
that an event falling into a certain bin is from elastic
scattering. Then, it is optimal to discard events with a low \Pel,
i.e. defining a decision function:
\begin{equation}
d(\vecQ) = \left\{ 
\begin{array}{l}
1\ \For\ \Pel(\vecQ) \geq \Pcut\\
0\ \For\ \Pel(\vecQ) < \Pcut\\
\end{array}\right.
\mylabel{eq:fuzzy4}
\end{equation}
where \Pcut(\mom,\Thcmsx) is to be chosen momentum and angle dependent.
Since the background originating from the carbon has been measured separately,
we can now quantify both the background reduction as well as the loss of
elastic events as a function of \Pcut. In \Fig{fig:fuzzy1}~(e) we show a
typical example. However, this procedure is done for every bin in
$(\mom,\Thcmsx)$ and every data sample individually.  
One obtains a sizeable reduction of background events (C) for minor
losses of elastic scattering events (H). We decided to sacrifice 3\%
of the elastic scattering events, allowing for a reduction of the
background of about 60\%. A typical contour of accepted events is
shown as a dashed line in  \Fig{fig:fuzzy1}~(d).
The events with $\Probpa \approx 0$ are populated mainly by events
where the time information of the scintillator bars is ambiguous due
to multiple hits. This occurs for some elastic events as well, so
that this class had to be kept. 

In order to determine \Pcut(\mom,\Thcmsx) with high {\em statistical}
precision we determined its value for 0.3~\GeV\ wide momentum  
and 10\grad wide \Thcmsx-bins. Its value is typically around 0.5 and 
smoothly varying with the beam momentum and \cms\ scattering angle.
When applying the cut to the final data, its value has been
interpolated for the correct momentum and angle.

\begin{figure}
\begin{center}
\includegraphics[width=\figwidth]{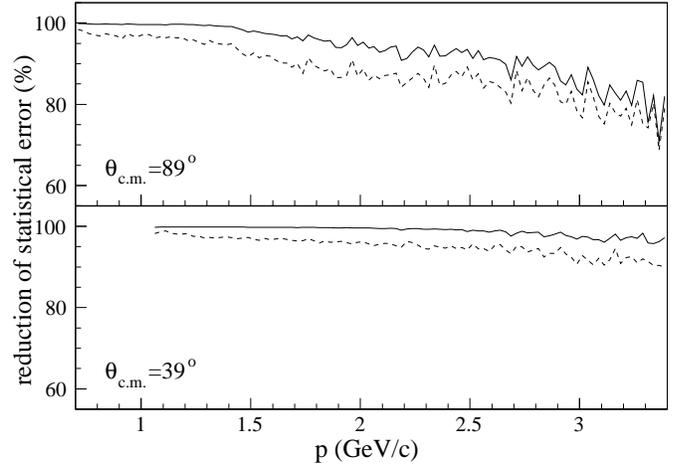}
\end{center}
\caption{Relative reduction of the statistical error from carbon
subtraction when applying the cut from the fuzzy-filter approach
(dashed line) and the reduction on the total uncertainty (solid line)
when all other sources of errors are included.  
}
\mylabel{fig:fuzzyErr}
\end{figure}

To summarize: by transferring the detector information -- by methods
adopted from fuzzy-logic -- to two variables with values between 0 and
1, any event can be mapped
onto the unit square. Since the data are binned we group the data in
400 classes according to their \Probpa and \Probpb values. Finally we
select a certain subset of these classes by requiring that the
respective $\Pel$ is above a certain threshold \Pcut , chosen to
result in only a small, well-defined loss of elastic events. 
This reduces the number of events, \Npp{\CH}{}\Arg\ and
\Npp{\C}{}\Arg, entering \Eq{eq:csub} to correct for the
contribution arising from carbon-contents of the target
(\Sec{carbonsubtract}). Since less events are subtracted
(\Figs{fig:cContribution}~(b) and 
\ref{fig:fuzzyAlpha}) the corresponding 
statistical uncertainty is considerably reduced 
(cf. \Fig{fig:fuzzyErr}), most notably at high momenta and large
\cms\ scattering angles where elastic cross sections are smallest. 

\begin{figure*}
\begin{center}
\includegraphics[width=\widefigwidth]{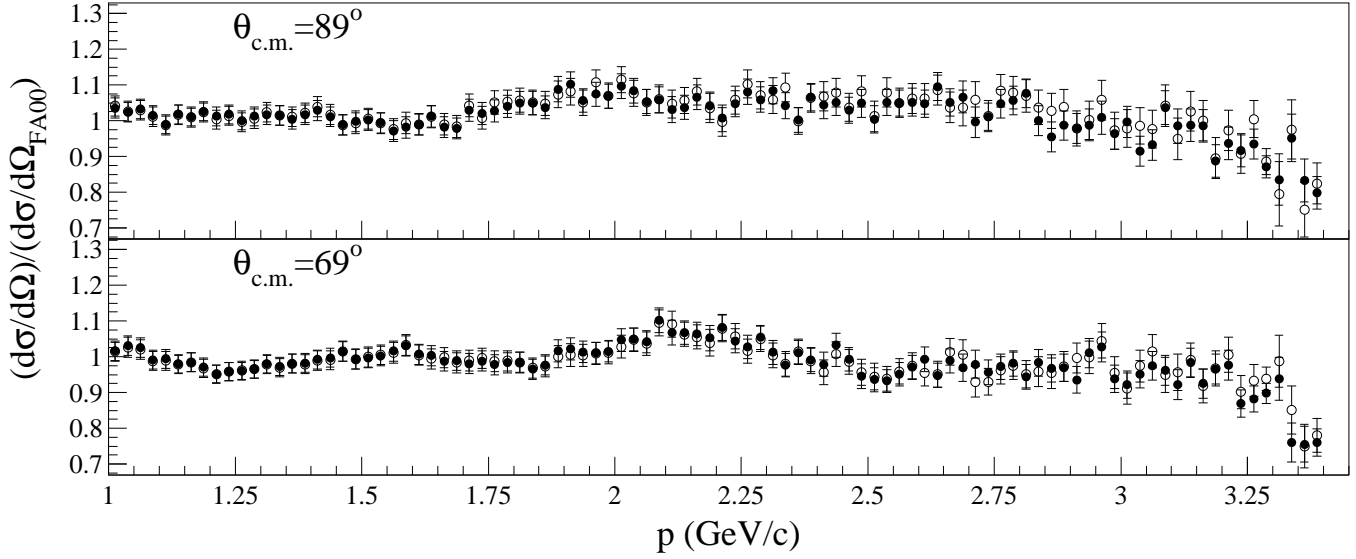}
\end{center}
\caption{Comparison of two excitation functions with (closed) and
without (open symbols) using the fuzzy-filter approach for background
subtraction. To remove the main energy dependence all cross sections have been
divided by the predictions of the PSA-solution FA00 of
\cite{Arndt:2000xc}.
}
\mylabel{fig:exciteFuzzy}
\end{figure*}

Since the resulting loss of elastic events is independent of beam
momentum and angle and its absolute value can be determined with 0.5\%
precision (e.g. 3\% $\pm$ 0.5\%) this is {\em not} offset by a large
increase of the {\em systematic} uncertainty. To this end a comparison
of the final cross sections with and without pre-selection of events using
fuzzy-logic (\Fig{fig:exciteFuzzy}) shows {\em no} systematic differences.
The precise value 0.5 of the parameter $g$ introduced in
\Eq{eq:fuzzy2} is not important, however, values close to $0$ or $1$
are to be avoided, since either the elastic events are concentrated in
only a few classes or spread out over all classes. Then \Pel\
can not be determined with sufficient precision for each class and the
net effect on background reduction is diminished. 

\subsection{Detection Efficiency}\mylabel{efficiency}

The probability $\eta(\mom, \Thcmsx)$ that an elastic pp-scattering event leads
to a valid trigger and survives all offline cuts was determined
by Monte-Carlo simulations.
It uses the available information on the vertex-distribution, beam
position and angles and accounts for the electronic thresholds and
trigger conditions. All simulated events are then analyzed by the same
software as used for the experimental data, applying the same
reconstruction algorithms and cuts. The fraction of events passing
these cuts $\eta(\mom, \Thcmsx)$, is fairly constant between 0.94 and
0.96 and shows no dramatic momentum dependence. 
The efficiency is smallest for the lowest momenta and
forward scattering and for the highest momenta and symmetric
scattering, i.e. $\Thlabi{1} \approx 
\Thlabi{2}$ or $\Thcmsx \approx$ 90\grad.
The main reduction is due to secondary reactions of the ejectiles
in the beam-pipe or the detector, so that the information on one or
both protons is sufficiently distorted to be removed on the
trigger-level or by offline-cuts.
The loss of 4 to 6\perc\ is consistent with the expectation derived
from the known total hadronic cross-sections and the thickness and
composition of matter in the detector set-up.
For application, $\eta(\mom,
\Thcmsx)$ is approximated by a polynomial expansion in
cos$\Thcmsx$ with momentum dependent coefficients. 
%
\subsection{Detector Acceptance}\mylabel{acceptance}
To account for the acceptance limits of the EDDA detector two
software cuts, on \Thcmsx\ and \myphi , are applied in the analysis:
The limits of the acceptance in the c.m. scattering angle \Thcmsx\ is
given by the requirement that both protons are detected in the
half-ring layer of the EDDA-detector. The minimum scattering angle
accepted for the final 
data (shown as the solid line in \Fig{fig:cContribution}) was selected
large enough to rule out elastic events to be lost due
to the extended beam-target overlap or small-angle scattering of the
ejectiles in the beampipe and the detector.

Although the EDDA detector yields full azimuthal coverage a cut on
\myphi\ is applied as well: For \myphi\ close to
90\grad\ or 270\gradx, events are lost on the trigger level when both
particles hit the same -- left or right -- half of the ring layer. 
In addition the angular resolution achieved with
the semi-rings is reduced at these angles, since close to the
readout large variations in the light collection
efficiency occur. Thus, we artificially reduced the azimuthal acceptance by 
selecting only events with the left particle emitted at
$\left|\myphi\right|\leq 76\grad$ and accounted for this reduced
acceptance with a factor $180/152 = 1.184 \pm 0.007$ in the
cross-section calculation.

%
\subsection{Normalization}\mylabel{lumi}

%
In internal target experiments the luminosity depends on the overlap
of the target density distribution with the transverse profile of the
stored beam. During acceleration both the horizontal beam position and
the emittance -- determining the beam width -- vary 
considerably (cf. \Fig{fig:vertex}) and cannot be determined, e.g. from
elastic scattering data (cf. \Sec{vertex}), with sufficient precision.   
Therefore, two electromagnetic processes with known dependence on the instantaneous
luminosity {\em and} beam energy were recorded concurrently with the
elastic scattering data. These are the current of secondary electrons (SEM)
emanating from the fiber target which is a function of the energy
deposit by the beam, 
and the rate of elastically scattered electrons, so-called
$\delta$-electrons, in PIN-diodes placed at 40\grad\ behind thin
windows in the beam pipe (cf. \Fig{fig:setuplumi}). These two methods are
referred to as SEM- and PIN-monitors. 

As it will be outlined in the following sections, both are not
accurate enough to allow an \EM{absolute} 
luminosity measurement.
However, the main uncertainties enter as multiplicative factors which
are independent of beam-momentum and can be eliminated by an absolute
normalization to high-precision reference data \cite{simon93} at \EM{one} beam
momentum (\pref~=~1.455~GeV/c). The change of the luminosity with respect to
this reference momentum is provided by the SEM- and PIN-monitors with
high accuracy, and will be referred to as \EM{relative} luminosity
determination.    

Furthermore, both monitors are sensitive to the electron contents of the target
only and measure $L_e$, i.e. the luminosity with respect to proton-electron
interactions. To relate this to the Luminosity $L_H$ for proton-proton
interactions the hydrogen \ahden\ and electron $\myrho_e$ densities
in the target enter, viz
\begin{equation}
L_H = \frac{\ahden(p)}{\rho_e} L_e = \rhden(p)
\cdot\underbrace{\frac{\hdenn}{\myrho_e} L_e}_{\makebox[0mm]{\LSEM\ {\rm or} \LPIN}},
\mylabel{eq:lumih}
\end{equation}
where \hdenn\ is the hydrogen density of an unused \CH-target.
The hydrogen density \ahden\ is the weighted average over the region
of the target sampled by the COSY-beam and therefore momentum
dependent. Its value will gradually decrease due to
radiation damage in the course of the experiment and will be discussed
in \Sec{radam}.
For convenience the factor \hdenn/$\myrho_e$ has been included in
\LSEM\ and \LPIN, so that it is fixed by the absolute normalization
and the factor $\rhden(p)$, common to both monitors, is deduced
separately. A corresponding, but much smaller change in the electron density will also
be taken into account.

\subsubsection{Secondary Electron Monitor (SEM)}
The SEM rate is proportional to the average energy $\overline{\Delta
E}$ {\it deposited} in the 
target fiber \cite{sternglass57}, and therfore to the restricted
energy loss rate \REL\
\cite{pdg98}, which takes into account that energetic knock-on 
electrons above some energy $T_{cut}$ will escape from the target and
not contribute to the energy deposit. We fix the cut-off energy  to
9~keV by requiring that the effective range of electrons equals half the target
thickness. However, any choice within a factor of 5 yields
the same {\em energy dependence} within 0.6\%, but not the same
absolute value of the restricted energy loss.
Thus, we expect the current \ISEM\ of emitted secondary electrons, as
recorded by a sensitive amperemeter \cite{scheid94}, to be given by
\begin{equation}
\ISEM = \kSEM\cdot\LSEM(\mom)\cdot\RELp.
\mylabel{eq:sem}
\end{equation}
The current is carried by the electrons emitted from the target
surface, and is dominated by electrons of very low
energy, produced by thermalization of the deposited energy. Hence, the 
proportionality factor \kSEM\ should not depend on the precise initial
distribution of the electron energies and thus the beam momentum --
it will, however,  be sensitive to the structure of the surface.
Any local damage of the aluminum-coating of the fiber and a resulting
possible charge-buildup would make \kSEM\ a function of the position
$x$ along the fiber. 
Since the proton beam moves along the target horizontally during
acceleration, this would result in an indirect
dependence of \kSEM\ on \mom. 
Such an effect has been ruled out by comparing \LSEM\ to the elastic
\pp-scattering rate when steering the beam across the target
horizontally during the flattop (cf. \Fig{fig:vertex}), so that $p$ is fixed
and only $x$ was varied. 
 
\subsubsection{$\delta$-Electron (PIN) Monitor}
\begin{figure}\SidecaptionBeg 
\includegraphics[width=5cm]{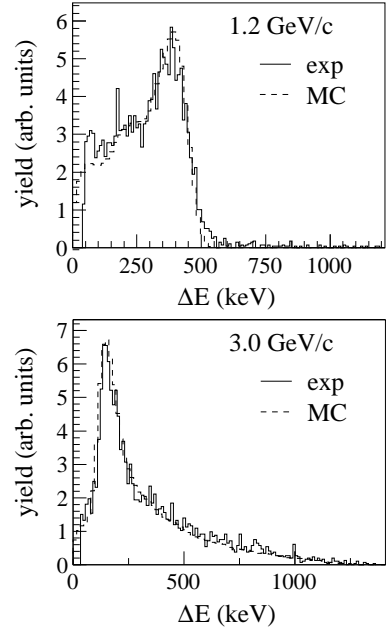}
\SidecaptionEnd
\caption{Comparison of the energy deposit of $\delta$-electrons in the
PIN diodes with MC-simulations.}
\mylabel{fig:pinmc}
\end{figure}

Elastically scattered electrons are detected at 40\grad by 500~$\mu$m
thick PIN diodes with an area of $10\times 10$~mm$^2$. They are mounted
in pockets at $\myphi = 0\grad$ and 180\grad in the beampipe, behind thin
(250~$\mu$m) aluminum windows. The electrons are
produced according to the well-known Rosenbluth cross section
\cite{Rosenbluth1950,kallen64E} with the simplification, that due 
to the small momentum transfers ($\approx$1~MeV/c) at COSY-energies
the form-factors are unity. 
Detection at 40\grad\ offers both statistical precision (
${\rm d}\sigma/{{\rm d}\Omega_{\rm lab}} \approx 200$~mb/sr) and 
electrons of sufficiently large energies (0.5-1.5~MeV) to be observed
with high efficiency.

The luminosity is determined in two steps: experimentally the rate of
scattered electrons $N_e$ is deduced from the singles rate in the
PIN-diodes $N_{\rm PIN}$, corrected for the rate of
hadrons $N_h$.
\begin{equation}
N_e = N_{\rm PIN} - N_h
\mylabel{eq:Ne}
\end{equation}
Since all electrons are stopped within the PIN-diodes or their mounting,
the rate of hadrons is determined by looking for coincident hits in
{\em both} bars and semi-rings of the outer detector layer
(cf. \Fig{fig:detector}) positioned behind the respective PIN-diode when
viewed from the target. We found that 4-6\% of all PIN-triggers had to
be attributed to hadrons, with the fraction linearly increasing with
beam momentum. To estimate the number of hadrons escaping detection in
the outer layer because of secondary reactions or stopping, a
telescope was mounted in place of a PIN-diode in an independent measurement.
The telescope comprised a 5x5~mm$^2$ PIN-diode followed by a
Cu-absorber, thick enough (0.7~mm) to stop all electrons, and a
$10\times 10$~mm$^2$ PIN-diode to detect the hadrons. It could be shown that
that 0.5\% of all PIN triggers not vetoed by the outer-detector layers
are due to hadrons \cite{hueskes97} and that this fraction is not
beam-momentum dependent.

Secondly, the number of electrons has to be related by a Monte-Carlo
simulation to the luminosity (cf. \Eq{eq:lumih})
\begin{equation}
\LPIN(\mom) = \frac{\rho_0}{\rho_e} N_e(\mom) \cdot l_{\rm MC}(\mom) 
\mylabel{eq:pin}
\end{equation}
where $\rho_0/\rho_e$ is the ratio of free protons to electrons in the
target and $l_{\rm MC}$ is the average luminosity in p-e elastic
scattering needed to produce
one detected electron. The latter is given by MC-integration of the  
Rosenbluth cross-section, taking into account the exact geometry of
the PIN-diodes, all materials used for mounting and shielding, the
COSY-beam direction, location and width (\Sec{vertex}) and the
detection thresholds determined 
experimentally. The interaction of the 
the electrons were modeled using the electron-gamma shower code
EGS4 \cite{EGS4}. 
In \Fig{fig:pinmc} the shape of the the experimental energy-loss
spectrum is compared to MC-simulations, showing the transition from
stopped electrons at lower energies to a typical Landau-like 
energy-loss distribution at higher energies. Note, that 
-- apart from normalizing to the same number of counts -- 
no parameters have been adjusted using the experimental data. The absolute
energy scale was fixed by calibration with conversion-electrons from a
$^{175}$Hf-source.  
The main source of systematic errors in $l_{\rm MC}$ \cite{gh01} are
uncertainties in 
\begin{enumerate} 
\item the target location along the beam, 
\item the horizontal beam position, 
\item the thickness of the aluminum window 
\item and the precise size of the depleted area of the 
PIN-diodes. 
\end{enumerate}
By taking the arithmetic mean of luminosity determined
with the PIN-diodes mounted to the left and right of the beam, the second
errors cancel and the last errors do not contribute when looking
only at changes of $l_{\rm MC}$ with beam momentum. 
The resulting systematic error of $\LPIN(\mom)/\LPIN(\pref)$ 
increases with $|\mom-\pref|$ and stays below 2.9\% (1\%) for beam
momenta below (above) \pref = 1.455~GeV/c. 

\begin{figure}[t]
\begin{center}
\includegraphics[width=\figwidth]{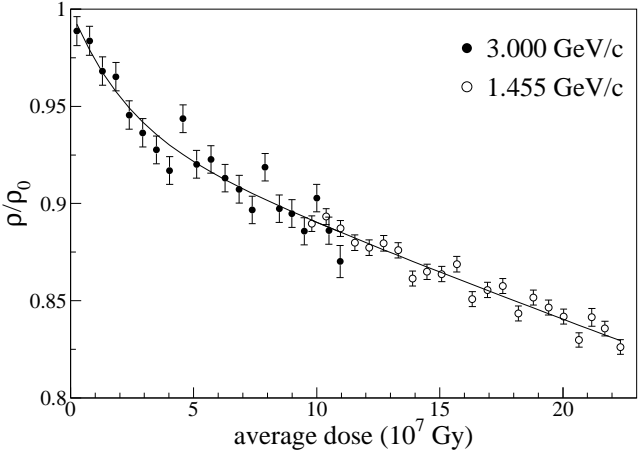}
\end{center}
\caption{Relative change of the hydrogen contents of \CH-targets as a
function of the accumulated dose as determined from measurements at
fixed energies. The solid line is a parameterization as outlined in the
text (from \cite{Rohdjess:2004a}).}
\mylabel{fig:hLoss}
\end{figure}

In principle an \absolute\ luminosity could be deduced, but its
precision hinges on the
knowledge of the ratio of electrons to protons in the \CH--target.
The naive estimate of $\rho_0/\rho_e$ being four is changed by the
aluminum coating ($\approx$20$~\mu$g/cm$^2$) to about
4.7. In addition, radiation damage (see \Sec{radam}) reduces the number
of hydrogen in the target, as outlined in the next section.
The absolute normalization to reference data agrees within 5-10\% with 
our estimates of $\ahden/\rho_e$, consistent with the errors due to
uncertainties of the aluminum thickness, the size of the depleted area
of the PIN diodes and the accumulated dose of a specific target.

\subsubsection{Correction for Radiation Damage of the Target}\mylabel{radam}
The hydrogen density of the \CH-target decreases during the course of
the experiment. This is due to radiation-induced 
cross-linking of the polypropylene polymers by replacing two C-H
bonds by a C-C bond and a H$_2$ molecule emanating
from the target. In comparison, losses due to hadronic interactions are
negligible. The relative change of the hydrogen density \hden, and to
a lesser extent of the electron density, with
respect to an undamaged target \hdenn\ will be a function of the
acquired dose. 
The dose is not deposited uniformly along the target,
but reflects the density distribution of the beam over the course of many
COSY machine cycles (cf. \Fig{fig:vertex}).
Finally, during the acceleration target regions with different
hydrogen density will be sampled at different momenta which needs to
be corrected for. 

To this end the history of each target with respect to the acquired
dose has to be closely monitored. The rate at which dose is acquired
is related to the instantaneous luminosity $L_H(t)$, the specific
energy loss of beam 
particles in the target and the normalized horizontal beam profile
$P(x,t)$ viz 
\begin{equation}
\frac{\dD}{\dt}(x,t) = 
\RELN\!\!\!\!\!\!\!\!\!\!\!\!\!\!\!\!(p(t)) \ \ \ 
\ \frac{k\,L_H(t)\,P(x,t)}{\vrhdennt},
\mylabel{eq:hloss1}
\end{equation}
where the constant $k = 113.6\,{\rm Gy\,mm\,mb\,cm/ MeV}$ encompasses
all information on target material and dimensions (see 
\cite{Rohdjess:2004a} for details). Since $L_H$ is the luminosity
with reference to the hydrogen content of the target, the factor
\vrhdenn\ corrects for the average relative hydrogen contents sampled
by the beam.

\begin{figure}[t]
\begin{center}
\includegraphics[width=\figwidth]{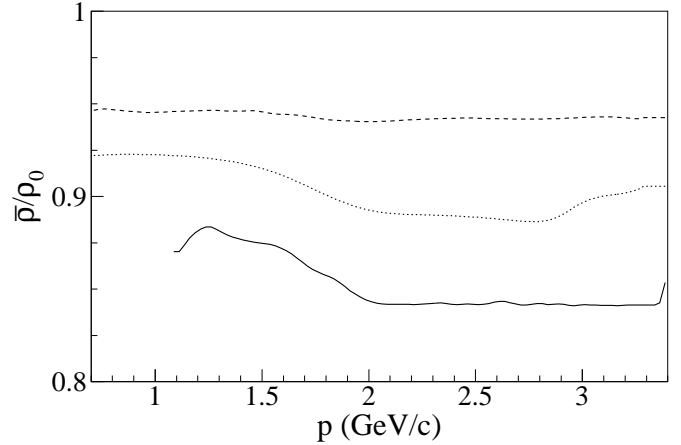}
\end{center}
\caption{Average relative hydrogen density as a function of momentum
for three samples corresponding to different setups of the COSY beam.
}
\mylabel{fig:hLossCorr}
\end{figure}

\begin{figure*}[t]
\begin{center}
\includegraphics[width=\widefigwidth]{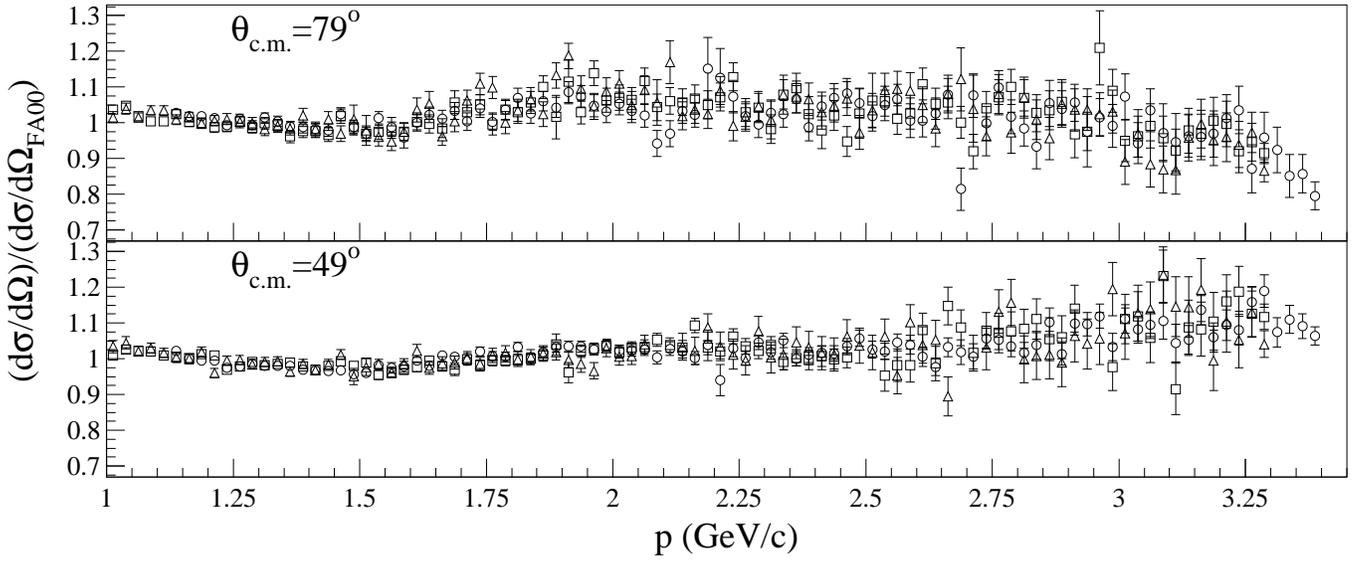}
\end{center}
\caption{Comparison of the differential cross section of
the three groups of samples (distinguished by the symbols {\Large
$\circ$}, $\Box$, and $\bigtriangleup$). Only statistical errors are
shown.
To remove the main energy dependence all cross sections have been
divided by the PSA-prediction of \cite{Arndt:2000xc} (solution FA00).
}
\mylabel{fig:exciteCompSamp}
\end{figure*}

The relative change of the hydrogen content has been determined
experimentally, by observing the relative change in the yield from 
pp-elastic scattering with respect to that of inelastic p-carbon
reactions at constant momentum as a function of the acquired dose,
calculated from the measured luminosity \cite{Rohdjess:2004a}. The
result is shown in 
\Fig{fig:hLoss} together with a parameterization of the form
\begin{equation}
\rhdenn(D) = f \exp(-\lambda_1 D) + (1 - f)
\exp(-\lambda_2 D). 
\mylabel{eq:hloss2}
\end{equation}
The loss is described by a fast process with decay constant
$\lambda_1 = (4.4 \pm 1.3) \cdot 10^{-8}~\Gy^{-1}$ which reduces the
hydrogen density by at most $ f = (5.8 \pm 0.7)~\%$, and a slow
process with decay constant $\lambda_2 = (5.67 \pm 0.28) \cdot
10^{-10}~\Gy^{-1}$, with an additional scale uncertainty of
$\lambda_{1,2}$ of 11\%.

For each target the luminosity was recorded for the time of exposure
to the COSY-beam and the beam profile $P(x,t)$ is known from the
reconstructed vertex distribution (cf. \Sec{vertex}) and the dose
can be calculated with \Eq{eq:hloss1} in a first approximation by assuming
$\vrhdenn=1$ to obtain the dose
distribution. Using the hydrogen density as given by \Eq{eq:hloss2}
the dose calculation can be iterated until self-consistency is
reached.

Finally, the hydrogen density as a function of the beam momentum
during the acceleration, averaged over the beam profile and the total
time of the measurement is calculated for each data sample. Three
examples, corresponding to different setups of the COSY-beam are
shown.
It turned out that for about one third of the data this correction is
not momentum dependent (dashed line in \Fig{fig:hLossCorr}), because
the change of beam position was small 
compared to the width of the beam. For other samples the correction
amounts to typically 5-15\%.

The shape of the correction factor only depends on the setup of the
COSY beam. The data samples have been taken with three different
settings, each one comprising roughly one third of the data. 
In \Fig{fig:exciteCompSamp} excitation functions obtained with these
three beam setups are compared at two c.m. scattering angles, showing
perfect agreement within statistics. When these cross sections are integrated
over the full detector acceptance of the detector, they agree much better
than the  2.5\% error corridor, as given by the relative
luminosity determination.

Note, that the PIN-monitor is affected indirectly by changes in
electron density $\rho_e$ of the target (cf. \Eq{eq:pin}). However,
most of the electrons are attached to the carbon and aluminum nuclei
and the applied correction is smaller by a factor of about 4.8. 
The SEM monitor is not affected by cross linking, since the SEM-yield is
determined by the electron density in the aluminum coating of the
target.

\subsubsection{Absolute Normalization}\mylabel{lumiAbs}

The SEM and PIN monitors record the change of the luminosity with beam
momentum very accurately. The absolute value is 
obtained by normalizing the pp-elastic scattering angular distribution
at {\it one} projectile momentum bin \pref\ to precise reference data
of differential cross sections. We use the precision measurement at
p$_{\rm ref}$ = 1.455\GeVc\ (T$_{\rm 
ref}$ = 793\MeV) by Simon et al.\ \cite{simon93}, where the absolute
normalization uncertainty is given as $\leq$ 1\percx.  
By numerically integrating the data points of \cite{simon93} we
arrive at a reference cross section
\begin{equation} 
	\begin{array}{rcl}
	\displaystyle \sigma_{\rm ref} & = & \displaystyle
	2\pi\int_{39.84\grad}^{88.51\grad}
	{\rm d}\Thcmsx\,\sin\Thcmsx
	\dsdocd{\rm ref}(\Thcmsx)\\\\
	 & = & \displaystyle 11.16 \pm 0.02^{\rm stat.} \pm
	0.11^{\rm syst.}~{\rm mb}.
	\end{array}
\mylabel{eq:absNorm}
\end{equation}
to be used for the final absolute normalization. By numerically
integrating the angular distribution at \pref\ of {\em this experiment}
over the same angular range, the values of \kSEM\ and $\rho_0/\rho_e$
(\Eqs{eq:sem}{eq:pin}) are adjusted to yield the same integrated
elastic scattering cross section $\sigma_{\rm ref}$.

The absolute normalization receives an additional statistical error
from the statistical precision of the angular distribution of our data
samples at the normalization momentum \pref. The procedure to minimize
the influence of this error on the final result when combining all
data samples is described in \Sec{consistency}.

\begin{figure}
\begin{center}
\includegraphics[width=\figwidth]{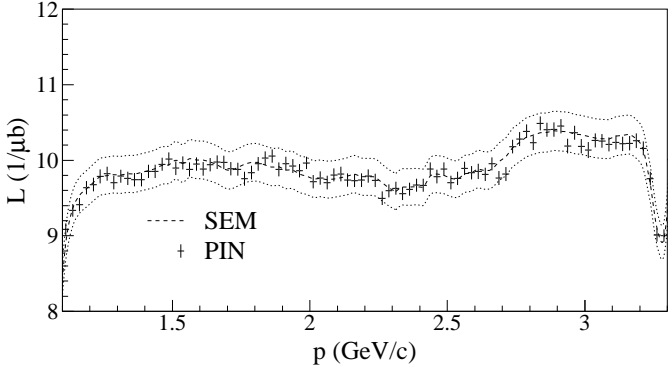}
\end{center}
\caption{Time-integrated luminosity for one out of 17 data samples from the
two luminosity monitors. The dashed line shows a $\pm$2.5\% band
around \LSEM\ to test the consistency to \LPIN. The absolute
normalization scale has been fixed at \pref~=~1455~\MeVc. Note the
suppressed zero on the vertical axis.}
\mylabel{fig:lumiAbs}
\end{figure}

\begin{figure}
\begin{center}
\includegraphics[width=\figwidth]{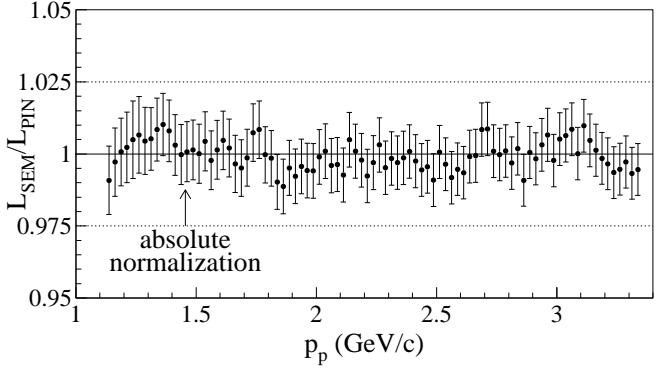}
\end{center}
\caption{Comparison of the SEM and PIN luminosity monitors averaged
over all data samples. The errors are dominated by the uncertainty in 
$l_{\rm MC}$.}
\mylabel{fig:complumi}
\end{figure}

\subsubsection{Comparison of Luminosity Monitors}\mylabel{lumiComp}
In \Fig{fig:lumiAbs} the result for the integrated luminosity from the
SEM- and PIN-monitors are compared for one typical data sample. The
agreement is typically much better than 2.5\%. The error on the
luminosity derived from e-p elastic scattering (PIN-monitor) is
dominated by systematic uncertainties from the
MC-simulations. Furthermore, below 1.1~\GeVc\ the $\delta$-electrons
at 40\grad\ are so low in energy, that they can no longer be detected
by the PIN-monitor. Therefore the luminosity obtained from the SEM-monitor
is used in the analysis and the PIN-monitor is taken as a consistency
check and to derive an error estimate on the {\em relative} luminosity
determination. When averaged over all data samples the two
luminosity monitors give consistent results (cf. \Fig{fig:complumi})
and deviations stay below 1.25\%\ within known statistical and
systematic errors. Since some samples show deviations as large as
2.5\%\ we use this value as an error estimate of the systematic {\em
relative} luminosity determination. This corridor is shown in
\Figs{fig:lumiAbs} and \ref{fig:complumi} as dotted lines.

\begin{figure}
\begin{center}
\includegraphics[width=\figwidth]{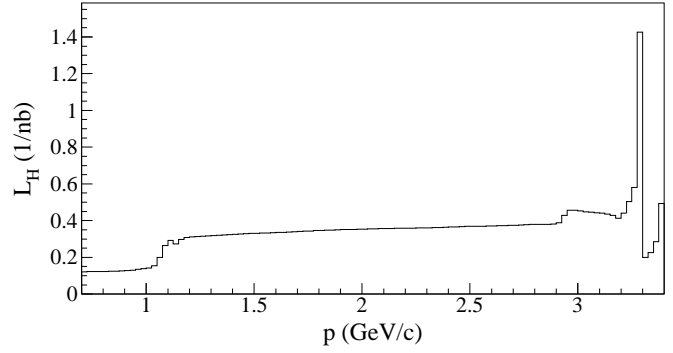}
\end{center}
\caption{Total integrated luminosity for all data.}
\mylabel{fig:lumiSum}
\end{figure}

The total absolute luminosity summed over all data samples is shown
in \Fig{fig:lumiSum}. The first data taking period covered the
momentum range 1.1-3.4~\GeVc\ and the second one 0.7-3.3\GeVc. The
spikes close to the maximum energy are due to the reduced ramping
speed in the transition to the flattop of the accelerator cycle. 
Note that the luminosity is more or less constant over the whole
momentum range (cf. \Figs{fig:lumiAbs} and \ref{fig:lumiSum}).
This shows that the loss of beam particles due to the fiber-target is 
compensated by the increase in beam current due to the rising
revolution frequency during acceleration. 
This is in line  with the finding \cite{hin89,hin92}, that the heating
of the stored beam by the fiber-target is counterbalanced by the
adiabatic damping of the betatron oscillations  
evinced as a near-constant beam-width $\delta x_v$ (cf. \Fig{vertex}).

\subsection{Beam Momentum Determination}\mylabel{momentum}
The beam momentum for each scattering event is determined from the
relative time $t$ of the event with respect to the start of the
COSY-cycle. This allows to calculate the nominal momentum $p(t)$ of the
COSY-beam based on the mathematical model used for programming the
function generators of all components of the COSY ring, especially the
cavity's RF and the dipole-currents. 

The beam momentum only depends on the revolution frequency $\nu$ and
the closed-orbit length $C$ of the stored beam:
\begin{equation}
p = m_p \beta\gamma c = m_p c \left( \left[ \frac{c}{C\ \fhf} \right]^2 -1\right)^{-\frac{1}{2}}
\mylabel{eq:beammom}
\end{equation}
Since COSY operates on the first harmonic the revolution frequency is
equal to \fhf. The RF-frequency was recorded during data taking, and
perfect agreement with its nominal value was found.
Time stamps were
recorded with the data every 2.5~ms using a high-precision 20~MHz clock
(relative accuracy of $10^{-6}$). Thus, based on the ramping speed of
1.15~\MeVc\ per ms, the momentum for individual events  
is known better than $\pm1.5$~\MeVc. 
The remaining uncertainty is due to the closed-orbit length $C$ and its
possible momentum dependence.
Upper limits on its deviation from the
ideal orbit length ($C_{\rm i} = 183.472$~m) were deduced from
measurements of the beam positions as a function of time at 29
locations around the ring. With 
the parameters of the COSY-lattice the maximal change in $C$
compatible with constraints 
from ion-optics was deduced. The corresponding relative momentum
deviation is at most
$3\cdot10^{-4}$, i.e. less than 1~\MeVc\ at 3.4~\GeVc\ 
\cite{engelhardt98}. A correction of $C$ with respect to $C_{\rm
i}$ turned out to be negligible.

\subsection{Dead-Time Correction}\mylabel{deadtime}
The dead time fraction of the data-acquisition system was typically
90\% and its instantaneous value had to be determined accurately. Using fast
(20~MHz) scalers we recorded both the number of events read-out $N_R$
and the number of all trigger $N_T$ (about 10~kHz) dead-time free.
The dead time fraction is then given by $\tau(t) = \left(1 - N_R/N_T\right)$ 
with statistical uncertainties of less than 0.5~\%. Systematic errors
were checked by taking data samples with data rates varying by orders
of magnitude and turned out to be negligible.

\subsection{Combination of Data Samples and Consistency Checks} \mylabel{consistency}

All 17 data samples were analyzed separately, the {\em relative}
luminosity was fixed using the SEM-monitor and the {\em absolute cross
section} normalization with respect to \cite{simon93} as described in
\Sec{lumiAbs}. These data samples are distinguished by at least one of
the following: the time of the measurement, the \CH-target used,
COSY-beam parameters and trigger conditions. This allows to test for
systematic variations of the results with these parameters.
Let us introduce the shorthand-notations $\sigma_i$ and $\delta\sigma_i$
for the differential cross section of the i'th data sample and its error. 
The compatibility of the results $\left\{\sigma_{i}\right\}$ of $n$
data samples, i.e. $i=1,\ldots,n$, was tested by looking at the
individual contributions to the total $\chiz$ when minimizing
\begin{equation}
\chiz = \sum_{i=1}^{n} \sum_{j,k}
\frac{\left(\Noi\,\sigma_i (\theta_{j},\mom_k) -\overline{\sigma_{j,k}}\right)^2}
	{\Noi^2\,\delta\sigma^2(\theta_{j},\mom_k)}
\mylabel{eq:combSamp}
\end{equation}
by variation of the normalization factors \Noi\ with \Nor{1} fixed to
unity. Here, the mean cross section is then given by the weighted mean
\begin{equation}
\overline{\sigma_{j,k}} = \left. \sum_{i=1}^{n} 
\frac{\Noi\,\sigma_i (\theta_{j},\mom_k)}{\Noi^2\delta\sigma^2(\theta_{j},\mom_k)}
\right/ \sum_{i=1}^{n} \frac{1}{\Noi^2\delta\sigma^2(\theta_{j},\mom_k)}.
\mylabel{eq:combSamp2} 
\end{equation}
The \Noi\ were treated as free parameters, so that the statistical
accuracy $\delta\sigma^2(\theta_{j},\pref)$ at the normalization
momentum, dominating the absolute 
normalization error of the individual data samples, do not
contribute to the \chiz. The fitted values for the \Noi\ turn
out to be very close to unity, well within the uncertainty due to $\delta\sigma^2(\pref)$.

When all data samples were combined the total \chiz\ per degree of
freedom was 1.03. 
When testing the influence of certain
aspects of the analysis on the final results, samples with the same
conditions with respect to the aspect under study were combined
viz \Eq{eq:combSamp} and 
renormalized to \cite{simon93} in order to increase the statistical
accuracy of the consistency check. An example is shown
\Fig{fig:exciteCompSamp} where the dependence on the COSY-beam setup
is tested, showing consistent results within the statistical errors.
No systematic deviation of the results from different samples could be
determined. In addition cuts applied to the data have been varied
within reasonable limits to check for possible systematic
dependencies of the final result on details of the analysis. As an example,
the value of \kdc, the maximum allowed value for the
kinematic deficit may be increased by 3\grad without changing the
results significantly, although at the expense of increased
statistical error. The change is always much smaller than the
estimated systematic error and amounts to less than 1\% for the
majority of the data. 

\begin{figure*}[t]
\begin{center}
\includegraphics[width=\widefigwidth]{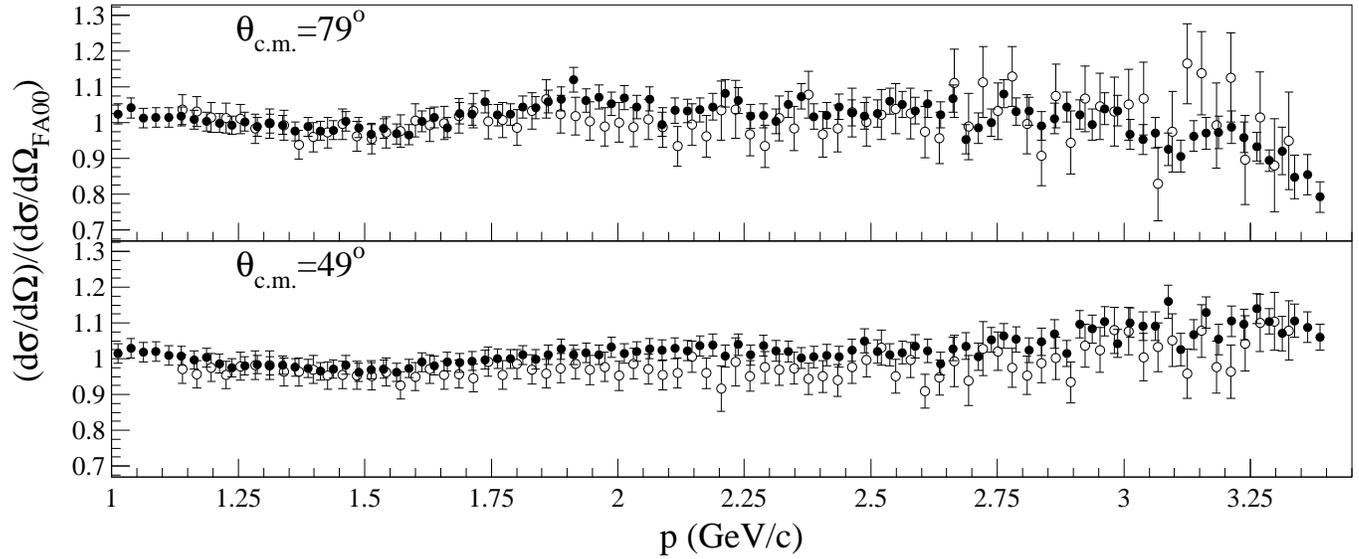}
\end{center}
\caption{Comparison of the differential cross section of
\cite{albers97} (open symbols) with this work (closed symbols) at two
angles. To remove the main energy dependence all cross sections have been
divided by the PSA-prediction of \cite{Arndt:2000xc} (solution FA00).
}
\mylabel{fig:exciteComp}
\end{figure*}

Final cross sections are obtained by combining all 17 samples with the
help of \Eq{eq:combSamp} and renormalizing them at \pref\ to \cite{simon93}.
These data comprise a total of $37\cdot 10^6$ pp elastic scattering
events.

\subsection{Errors and Uncertainties} \mylabel{error}
The calculation of the excitation functions receives error
contributions from all factors on the right hand side of
\Eq{eq:2A}. We distinguish five different kinds of errors:
\begin{enumerate}
\item statistical errors, due to the number of observed counts from
\CH- and C-targets as in \Eq{eq:csub} with a very small contribution
from the relative luminosity $\Lum$ of the carbon-sample. This error
ranges from 1\% at small \Thcms and small momenta to 7\% at \Thcmsx
$\approx 90$\grad and large momenta reflecting the functional dependence of \dsdoc.

\item systematic errors which are treated as uncorrelated between
data points adjacent in scattering angle or momentum. Here, artefacts
from the reconstruction of the scattering angles contribute. The
angular resolution of the EDDA detector is not entirely homogeneous,
but varies slightly with $Q_{12}$ (cf. \Eq{eq:4}). When the data are
binned in \Thcmsx\ an uncertainty of 2.4\%, usually {\em
anti-correlated} for neighboring bins, arises. Smaller contributions
to this uncertainty stems from the cut on the azimuthal angle $\phi$
(0.6\%) and the loss of events due to the cut on \Pel\ (0.5\%).

\item systematic errors associated to undetected elastic scattering
events and misidentified inelastic events. This uncertainty will have
a smooth dependence on scattering angle and beam momentum. The
detection efficiency, obtained by MC methods, contributes  with 1.5\%
while possible contamination with inelastic events account for an
uncertainty increasing with beam momentum and scattering angle from less
than 1\% to 5.5\%. These contributions are always smaller than the first
two errors combined. 

\item a {\em relative} normalization error, common to all data points
at the same momentum of 2.5\%, derived
from the maximum discrepancy observed in the relative luminosity
determination of the SEM and PIN monitors. The statistical error of
the SEM signal as well the uncertainty in determination of the
DAQ dead-time are negligible in comparison. Uncertainties in the
correction \vrhden, due to uncertainties in $\lambda_{1,2}$ of
\Eq{eq:hloss2} and the applied dose, enter only as the ratio to its
value at \pref, so that its contribution is less than 0.3\% for all
beam momenta.

\item an {\em absolute} normalization error common to all data of less
than 1.5\%, comprising
the uncertainty of  $\sigma_{\rm ref}$ of \Eq{eq:absNorm} and the
statistical errors of our data at \pref.
\end{enumerate}
In all figures presenting the data of this work, only the first two
errors are shown, i.e. the statistical error and an additional
2.5\% systematic error. All errors are listed in \cite{DataAccess}.

\section{Experimental Results}\mylabel{results}
The experimental results of the present elastic $pp$ scattering 
experiment are the unpolarized differential cross sections
\dsdoc{}\
as a function of the c.m.-scattering angle \Thcms\
and the laboratory momentum $p$ of the proton beam.
Each differential cross section refers to bins $\Delta\Thcmsx= 2\gradx$ and
$\Delta p=25$~MeV/c, which were chosen to have reasonable statistical precision
for each data point. 
We covered a beam momentum range of $0.7 - 3.4$~GeV/c
corresponding to laboratory kinetic energies $0.23 - 2.59$~GeV
and total c.m. energies $\sqrt{s}$ = $2.0 - 2.9$~GeV.
Since in pp elastic scattering the final state particles are
indistinguishable and cross sections are symmetric with respect to
$\Thcms =90$\gradx, data are only given for the 
c.m. angular range $34\gradx - 90\gradx$.
All 2888 data points are available online through the world-wide web
\cite{DataAccess}. They may be viewed as 28 excitation functions
(cf. \Fig{fig:excite}) at c.m. scattering angles between 34\grad\ and
90\grad\ or  108 angular distributions  (cf. \Fig{fig:angDis}) at 108
beam momenta between 0.7 and 3.4~GeV/c. Note, that due to kinematics
the angular acceptance is reduced at very small and very high beam momenta.

As compared to our previously published results \cite{albers97}, based
on a subset of the data, the
momentum range is increased from 1.1-3.3~GeV/c to 0.7-3.4~GeV/c and
the statistical precision increased by up to a factor of 3. In addition,
in the analysis of Ref.~\cite{albers97} the variation of the hydrogen
density along the target due to radiation damage was not corrected
for, this led to a systematic underestimation of the cross section
by about 5\% above 2~GeV/c, as displayed in \Fig{fig:exciteComp}.
Note, that the results from \cite{albers97} are superseded by
this work and should not be used anymore.

\subsection{Comparison to Other Data}\mylabel{comparison}
In \Fig{fig:exciteData} the available database on pp elastic scattering 
(cf. \cite{Arndt:1985vj,Arndt:1992kz,Arndt:1997if,Arndt:2000xc,SAID} and
references therein) is plotted for the same angles as for the data of
the present work in \Fig{fig:excite}. The benefit of a consistent
normalization, made possible by measuring during acceleration in an
internal experiment, as well as the improved statistical accuracy is
evident. Previous data scatter considerably 
around available phase-shift solutions (discussed in detail in  \Sec{PSA}) 
partly due to larger statistical errors, but mainly due to differences
in the absolute cross section normalization of the various
experiments. Most notably the data of Jenkins et al. \cite{jenkins80} from the
ZGS are lower by about 20\% ($\bigstar$,\Star\ in
\Fig{fig:exciteData}) and clearly disagree with our data as well as
those of \cite{kammerud71,williams72}. In \Fig{fig:angDisData} angular
distributions are compared at four momenta to data from other
experiments. The shapes of the differential cross section are consistent
within quoted uncertainties, however, again absolute normalizations
are at variance. To show the size of this scale difference we have
renormalized our data using the procedure described in \Sec{lumiAbs}
to match the normalization of other experiments which spanned a larger
range of beam momenta. The result is displayed in \Fig{fig:compnorm}
and shows that the data of Kammerud et al. \cite{kammerud71} are
consistently higher by 10~\%, the data of Williams et
al. \cite{williams72} agree, within the sizeable uncertainties, and
the data from Albrow et al. \cite{albrow70} are partly compatible with
our normalization. On the average these data tend to be {\em larger} at
momenta above about 1.5~GeV/c, with the exception of the data of
Jenkins et al. \cite{jenkins80} discussed earlier. The excitation
function of Garcon et al. \cite{garcon85} at $\Thcmsx \approx
90\grad$, measured by a  
similar technique, shows the opposite trend (\Fig{fig:garcon}) and
cross sections are {\em smaller} by up to 2 standard deviations.  

\begin{figure}
\begin{center}
\includegraphics[width=\figwidth]{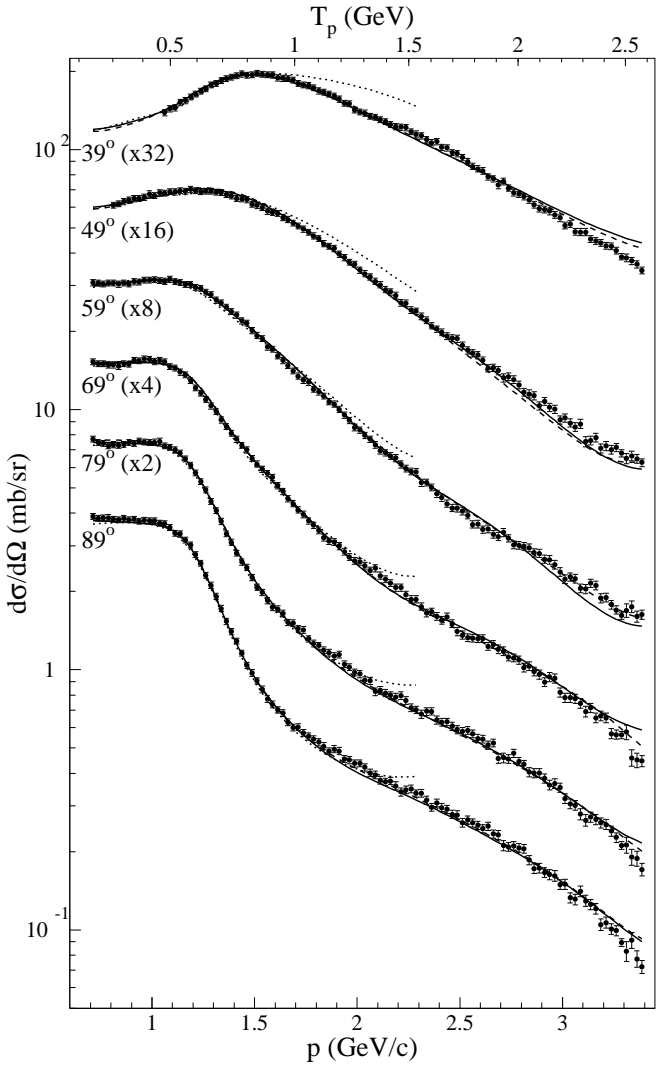}
\end{center}
\caption{Excitation functions of \dsdoc{}\ for six \cms\ scattering
angles in comparison to SAID solutions SM94 (dotted), SM97 (dashed)
and FA00 (solid line).
}
\mylabel{fig:excite}
\end{figure}

\begin{figure}
\begin{center}
\includegraphics[width=\figwidth]{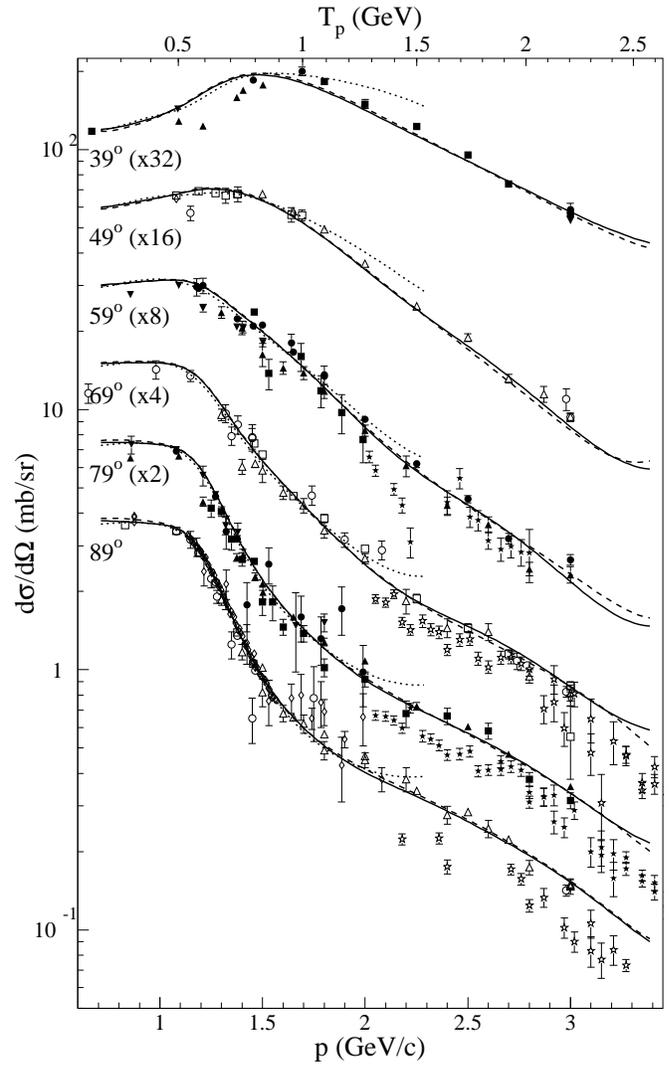}
\end{center}
\caption{Collection of published data (From the database of
\cite{Arndt:2000xc} without EDDA data of Albers et al. \protect\cite{albers97}) plotted as excitation functions at six angles
($\pm1\gradx$) in comparison to phase shift predictions as in
\Fig{fig:excite}.
}
\mylabel{fig:exciteData}
\end{figure}

\begin{figure}
\begin{center}
\includegraphics[width=\figwidth]{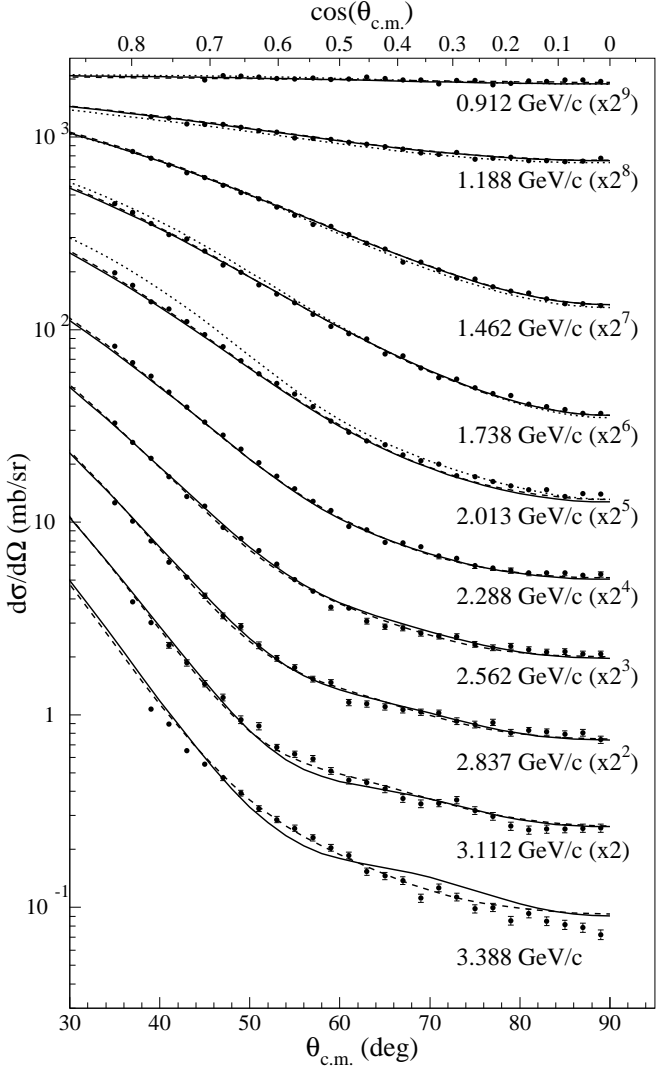}
\end{center}
\caption{Angular distributions of \dsdoc{}\ ({\Large$\bullet$}: data of this
work) for ten beam momenta $p$
in in comparison to SAID solutions (cf. \Fig{fig:excite}).
}
\mylabel{fig:angDis}
\end{figure}

\begin{figure}
\begin{center}
\includegraphics[width=\figwidth]{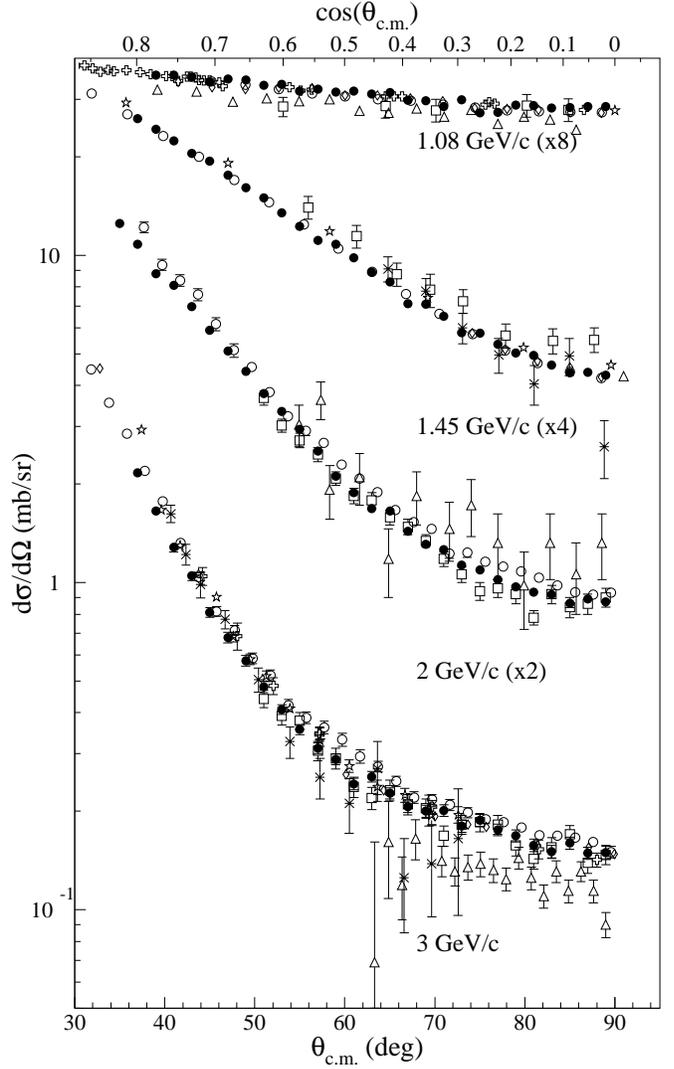}
\end{center}
\caption{Angular distributions at four beam momenta. Data of this
work ({\Large$\bullet$}) are compared to published data, where
the symbols {\Large\Circ}, $\square$, $\triangle$,
$\diamondsuit$, \Cross, \Star, and \Ast\ correspond to data from
\cite{simon93,albrow70,Abe:1975hs,simon96,Hoffmann:1988jw,Ottewell:1984hh}
at 1.08~GeV/c,
\cite{simon93,albrow70,garcon85,simon96,Dobrovolsky:1983wz,Barlett:1983ky,Ryan:1971rv}
at 1.45~GeV/c, \cite{shimizu82,kammerud71,williams72}  at 2~GeV/c and
\cite{kammerud71,williams72,jenkins80,Clyde:1966,Ankenbrandt:1968,Rust:1970wp,ambats74}
at 3~GeV/c.
}
\mylabel{fig:angDisData}
\end{figure}

\begin{figure}
\begin{center}
\includegraphics[width=\figwidth,bb=9 15 523 352]{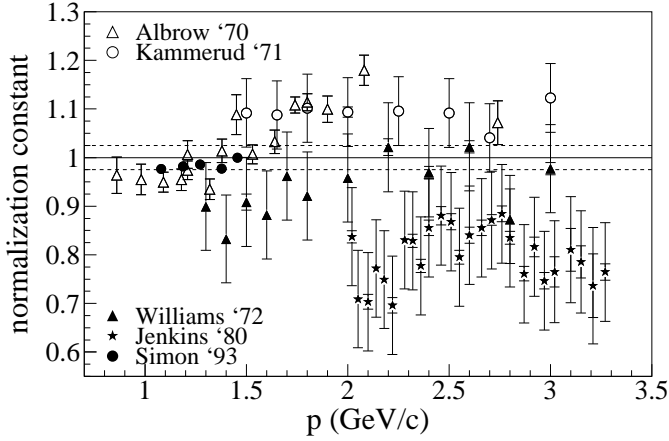}
\end{center}
\caption{Comparison of the absolute normalization constant relative to
\protect\cite{simon93} at $1.455$~\GeVc\ for angular distributions measured at 
CERN (Albrow et. al \protect\cite{albrow70}), 
the Argonne ZGS (Kammerud et. al \protect\cite{kammerud71}),
Rutherford (Williams et al. \protect\cite{williams72}),
Brookhaven (Jenkins et. al \protect\cite{jenkins80}), and
LAMPF (Simon et al. \protect\cite{simon93}). The smaller error bars show
statistical uncertainties only and the larger error bars contain the
normalization uncertainty added in quadrature. The dashed horizontal
line shows the relative normalization error of this experiment. 
}
\mylabel{fig:compnorm}
\end{figure}

\begin{figure}
\begin{center}
\includegraphics[width=\figwidth]{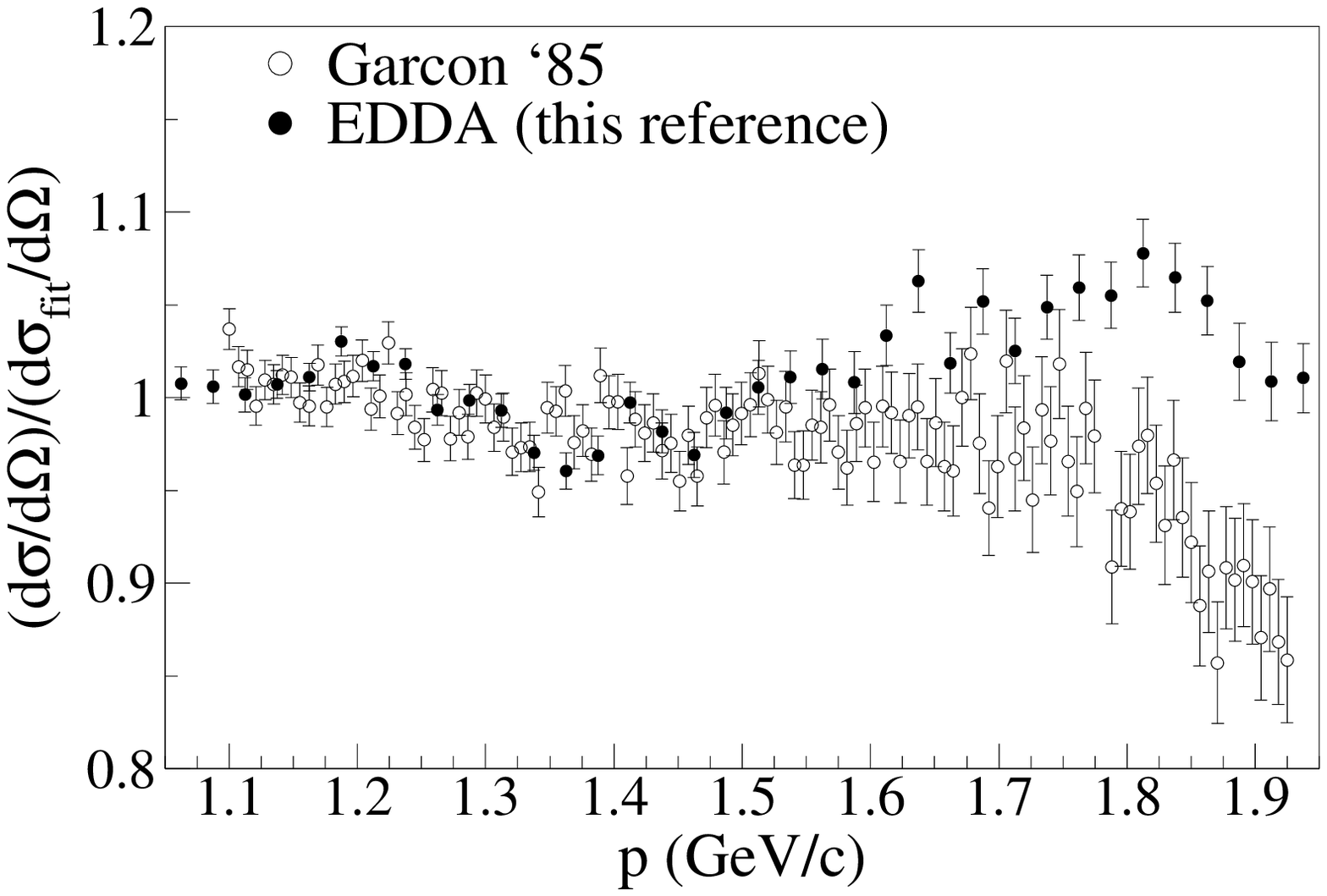}
\end{center}
\caption{Comparison to the excitation function of \protect\cite{garcon85}
at $\Thcms\approx 90\grad$. The EDDA date have been interpolated to
the exact angle as in \protect\cite{garcon85} and all cross sections have been
divided by \dsdoc{fit}, a smooth parameterization of the data.
}
\mylabel{fig:garcon}
\end{figure}

Note, that all 2888 cross section data obtained in  this experiment
share a common absolute normalization factor. In case data becomes
available in the future, with a superior method to obtain an absolute
normalization, our data could be renormalized by a common factor to be
determined along the same lines as described in \Sec{lumiAbs}. The
{\em relative} normalization uncertainty of 2.5~\% would not be affected.

\begin{figure}
\begin{center}
\includegraphics[width=\figwidth]{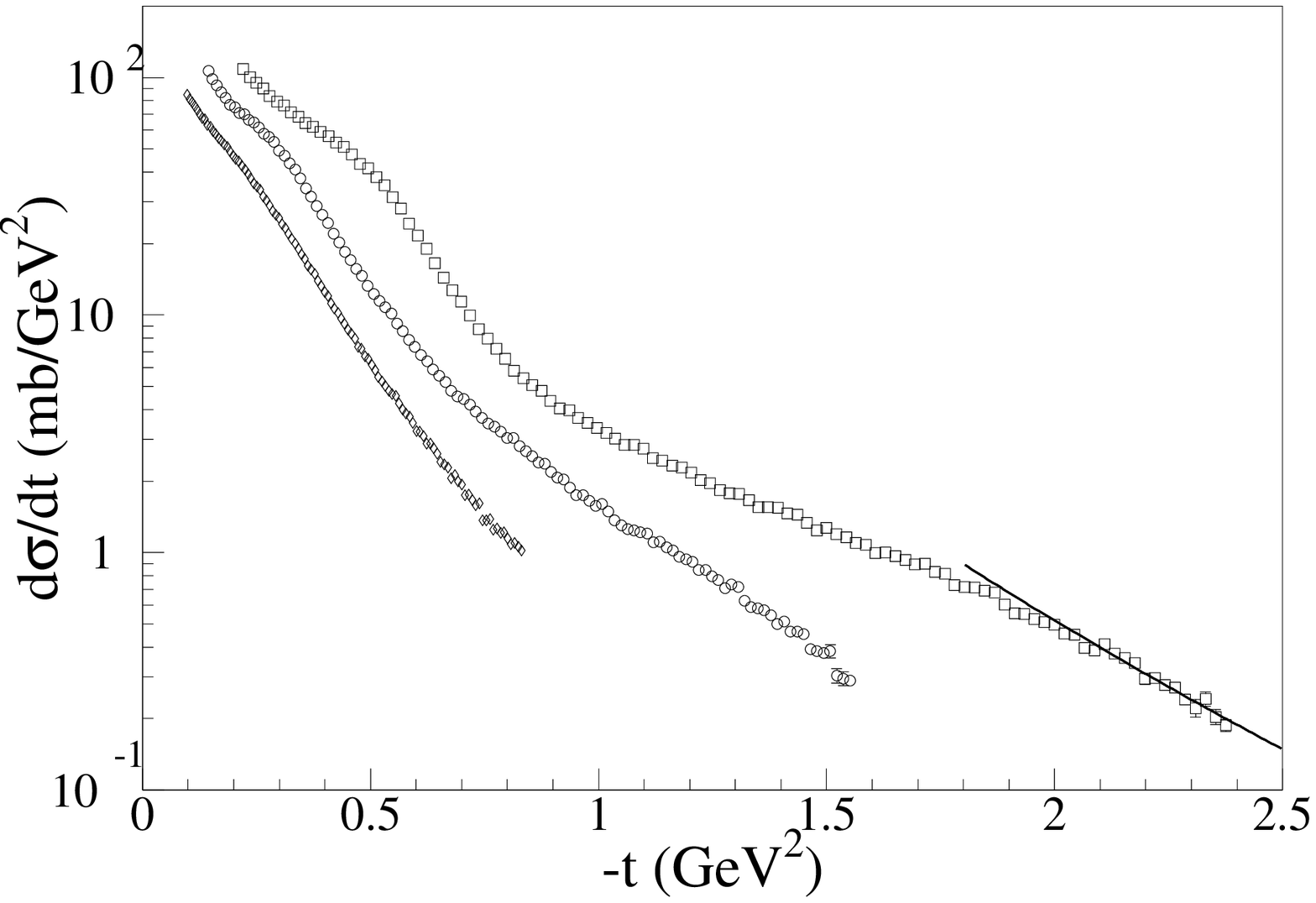}
\end{center}
\caption{Differential cross section \dsdt{}\ at 3 selected c.m. scattering
angles: 89\grad ($\square$), 69\grad ({\Large\Circ}), and 49\grad ($\diamondsuit$).
}
\mylabel{fig:exciteT}
\end{figure}

\section{Discussion}\mylabel{disc}
Theoretical models of the nucleon-nucleon interaction have been very
successful in describing elastic scattering data at energies where
cross sections of inelastic reactions are still small. In this domain
(\Tp\ below about 0.5~GeV) high-precision potentials, purely
phenomenological or based on the meson-exchange picture, are available
(for a review see Ref.~\cite{machleidt89}). More recently
effective field-theory has entered the stage and produced results of
comparable quality,
cf. Refs.~\cite{vanKolck:1999mw,Epelbaum:1999dj,Entem:2003ft} and
references therein. However, in the energy region between 0.5~GeV and a
few GeV little progress has been made in recent years. In the late
80s, the meson exchange models were extended to higher energies by
including inelastic reactions by coupling to intermediate $N\Delta$
and $\Delta\Delta$ channels \cite{Elster:1988pp,Elster:1988zu}. A
qualitative description of the $NN$-data below 1~GeV has been
achieved. However, at higher energies these models fail badly
and differential cross sections at large scattering angles are
strongly overestimated \cite{Eyser:2003zw}.
This may not be surprising since inelasticities at these energies will
no longer be driven predominantly by the $\Delta(1232)$ resonance. 
Work aiming to improve these models are currently under way \cite{Elster:2004}. 

On the high-energy side optical potential models and Regge-phenomenology
has been used to describe the gross features of the unpolarized
differential cross section \cite{kammerud71, jenkins80}. 
Dimensional scaling \cite{Brodsky:1973kr} predicts cross sections to
follow a $s^{-10}$ behavior at large momentum transfer. 
Our data are too low in momentum transfer for these models to
be strictly applicable. Nonetheless, in \Fig{fig:exciteT} we compare our
differential cross sections, plotted as \dsdt\ vs $-t$, to the
prediction of \cite{Brodsky:1973kr} (solid line, note that for 
\Thcmsx = 90\grad\ we have $s = 4m_p^2 -2t$), showing that we may be barely
touching the region of dimensional scaling, in
agreement with \cite{jenkins80}.

The immediate benefit of the high-precision data obtained in this
experiment is two-fold: first, its consistent normalization allow for
stringent tests of energy-dependent structures, as they could arise
from coupling to an intermediate resonant state, and secondly it will
further help to consolidate phase-shift parameters for the isotriplet
elastic NN-channel. 
 
\subsection{Upper Limits on Resonant Contributions}\mylabel{excitationfu}

All excitation functions show a smooth 
and rather structureless dependence on beam momentum.
No sharp energy-dependent structure is observed which could
be taken as evidence for a narrow resonance.
Therefore, upper limits for
the elasticities $\Gamma_{el}/\Gamma$
of hypothetical narrow resonances are deduced
from the smooth excitation functions.
To this end a Breit-Wigner resonance term is
introduced into the scattering matrix
element of the  partial wave
which is assumed to exhibit resonance behavior.
The interference between the Breit-Wigner resonance term and
the non-resonant amplitudes  determines the
size of a resonance excursion in the excitation functions.
The non-resonant amplitudes represent the
null-hypothesis.
In order to establish the null-hypothesis
the EDDA data are fitted by a
special energy dependent phase shift analysis
along Ref.\cite{Arndt:2000xc}.

In the test calculations the hypothetical resonance energy
$E_R$, total width $\Gamma$ and
resonance phase $\phi_R$ are varied systematically in the range
$E_R=2.2 - 2.8$~GeV, $\Gamma = 10 - 100$~MeV and 
$\phi_R = 0\grad - 360\grad$.
The hypothesis of the existence of a resonance
in a partial wave is tested by gradually
increasing the partial elastic width $\Gamma_{el}$
until the resonance is excluded within 99\% confidence
level by a \chiz-test based on the 
the differential cross sections and analyzing power data \cite{Altmeier:2000}.
For the unknown phase $\phi_R$ the
value giving the largest, and thus most conservative limit on
$\Gamma_{el}/\Gamma$ is chosen. 
Typical upper limits of $\Gamma_{el}/\Gamma$
for the five lowest uncoupled partial waves
are 0.08 for $^1$S$_0$, 0.04 for $^1$D$_2$,
0.10 for $^3$P$_0$,
0.03 for $^3$P$_1$ and 0.05 for $^3$F$_3$.  
For instance the $^1$S$_0$ dibaryon resonance 
predicted by Lomon et al. \cite{Gonzalez:1987gj} 
at $E_R = 2.7$~GeV with $\Gamma_{el}/\Gamma = 0.1$
and $\Gamma = 50$~MeV can safely be excluded.
The method is described in detail in a forthcoming paper \cite{Rohdjess:2004b}.

It should be noted in this context that
two broad resonant structures are well known at lower energies,
the $^1$D$_2$ resonance at $E_R \approx 2.15$~GeV 
and the  $^3$F$_3$ resonance at $E_R \approx 2.17$~GeV with widths of around
120~MeV. They appear as counterclockwise circles in the
Argand diagrams of the phase shift analysis \cite{Arndt:1992kz}
of the world data set.
Especially the 
excitation functions of the spin-dependent
total cross sections $\Delta \sigma_L$
and $\Delta \sigma_T$ measured at ZGS \cite{Auer:1978aa,bieler78}
show marked energy-dependent structures.
These  $^1$D$_2$ and $^3$F$_3$ resonances in the $pp$-system
can be interpreted conventionally by intermediate 
$^5$S$_2$ and $^5$P$_3$ $N\Delta$ states
\cite{bug84,Elster:1988zu,Gonzalez:1986pd,LaFrance:1993aq,Ueda:1982fz}.
These broad resonances show up as a rather rapid change in the shape of the
angular distribution around 1.25~GeV/c ($\sqrt{s}=2.15$~GeV)
where the cross section exhibit a steep ascend (descend)
at forward (backward) angles.

\subsection{Impact on phase shift analysis}\mylabel{PSA}
The excitation functions of the differential cross sections
are measured over a wide momentum range from 0.7 to 3.4~GeV/c.
They have a great impact on the energy-dependent
phase shift analysis since they represent a precise and consistent data set
of 2888 data points. 
Another important fact is that 
the unpolarized differential cross sections
fixes the global scale of all amplitudes and thus
enters repeatedly into the phase shift analysis through all spin
dependent cross sections, which are products of
spin-observables with the unpolarized cross section.

The dotted curves in  \Fig{fig:excite} are phase shift solutions SM94 of Arndt et al. \cite{arndt94d}
from summer 1994, i.e. before the first EDDA data were available.
The maximum kinetic energy of this phase shift solution was
1.6~GeV corresponding to a beam momentum of 2.36~GeV/c.
After the publication of the first EDDA data \cite{albers97}
the VPI group extended their energy dependent phase shift analysis
from 1.6~GeV (2.36~GeV/c) up to 2.5~GeV (3.3~GeV/c), with the solution
SM97.
Meanwhile an energy dependent phase shift solution FA00
is available up to 3.0~GeV (3.82~GeV/c)
laboratory kinetic  energy (beam momentum) \cite{Arndt:2000xc}. 
The phase shift solutions SM97 and FA00 are shown
as dashed and solid curves in \Figs{fig:excite} and \ref{fig:angDis} ,
respectively. 
Phase-shift predictions show an apparent oscillation both in angle and
beam momentum about the data of this work. This may be an
artefact of the current parameterization of the energy-dependence of
phase-shifts. Therefore, including the data of the present work will
not only slighly modify the phase-shifts to reflect the differences to
our previous results (\cite{albers97} and \Fig{fig:exciteComp}) but
may also allow to improve the ansatz for the variation of the
phase shift parameters with energy. In this process a better
description of the angular distribution at the higher momenta should
be attempted. 

\section{Summary}\mylabel{sum}

In a dedicated experiment, protons accelerated in a
synchrotron have been scattered of internal \CH-fiber targets {\em
during accelaration} at beam momenta between 0.7 and 3.4~GeV/c ($\Tp = 0.23 \ldots 2.59$~GeV). Elastic
scattering events have been identified by the EDDA-detector over a
wide angular range ($\Thcmsx = 34\grad \ldots 90\grad$). 
This experimental technique allows a precise monitoring of the
relative change in  luminosity with beam momentum which leads to a
consistent normalization of data taken at different momenta and is
therefore ideally suited to measure excitation functions with high
precision. The absolute cross section scale was fixed at the reference
momentum 1.455~GeV/c to high-precision data from LAMPF \cite{simon93}.
The average combined statistical and systematic error is 3.8\%, it
ranges from 3\% for small momenta and scattering angles to 10\% at high
momenta and $\Thcmsx = 90\grad$. The total {\em absolute} normalization
uncertainty of the complete data set is below 1.5\% with an
additional {\em relative} normalization uncertainty of 2.5\% common to
all data points at the same beam momentum.

The wealth of new data obtained in this experiments replaces the data
reported earlier \cite{albers97} and utilized improved analysis
methods to yield higher statistical and systematic precision, most
notably at higher energies. These data will further improve extracted
phase-shift parameters and provide an important normalization standard
for measurements of NN reactions. The data are available via
the world-wide web \cite{DataAccess}.

The excitation-functions obtained with steps of 25~MeV/c in
beam-momentum (or $\Delta\sqrt{s} \approx 8.5$~MeV) have been applied in 
tests for contributions of narrow resonances, like dibaryons,
coupling to the elastic channel. No evidence for any narrow
(10-100~MeV) structures have been found in the invariant mass range
$2.0\ldots2.8$~GeV covered by the experiment, ruling out a strong coupling
of such resonances -- if they exist -- to the elastic channel.

\Beginack
We thank the operating team of COSY for excellent beam 
support. This work was supported by the German BMBF, contracts 06BN664I(6) and 06HH852, 
and by the Forschungszentrum J\"ulich (Auftr\"age  411 268 03 and 41126903). 
\Endack

\printfigures


\begin{thebibliography}{76}
\expandafter\ifx\csname natexlab\endcsname\relax\def\natexlab#1{#1}\fi
\expandafter\ifx\csname bibnamefont\endcsname\relax
  \def\bibnamefont#1{#1}\fi
\expandafter\ifx\csname bibfnamefont\endcsname\relax
  \def\bibfnamefont#1{#1}\fi
\expandafter\ifx\csname citenamefont\endcsname\relax
  \def\citenamefont#1{#1}\fi
\expandafter\ifx\csname url\endcsname\relax
  \def\url#1{\texttt{#1}}\fi
\expandafter\ifx\csname urlprefix\endcsname\relax\def\urlprefix{URL }\fi
\providecommand{\bibinfo}[2]{#2}
\providecommand{\eprint}[2][]{{#2}}

\bibitem[{\citenamefont{{Lechanoine-LeLuc} and Lehar}(1993)}]{Lechanoine:1993}
\bibinfo{author}{\bibfnamefont{C.}~\bibnamefont{{Lechanoine-LeLuc}}}
  \bibnamefont{and} \bibinfo{author}{\bibfnamefont{F.}~\bibnamefont{Lehar}},
  \bibinfo{journal}{Rev. Mod. Phys.} \textbf{\bibinfo{volume}{65}}
  (\bibinfo{year}{1993}).

\bibitem[{\citenamefont{Haeberli et~al.}(1997)}]{Haeberli:1997}
\bibinfo{author}{\bibfnamefont{W.}~\bibnamefont{Haeberli}}
  \bibnamefont{et~al.}, \bibinfo{journal}{Phys. Rev.}
  \textbf{\bibinfo{volume}{C55}}, \bibinfo{pages}{597} (\bibinfo{year}{1997}).

\bibitem[{\citenamefont{von Przewoski et~al.}(1998)}]{vonPrzewoski:1998ye}
\bibinfo{author}{\bibfnamefont{B.}~\bibnamefont{von Przewoski}}
  \bibnamefont{et~al.}, \bibinfo{journal}{Phys. Rev.}
  \textbf{\bibinfo{volume}{C58}}, \bibinfo{pages}{1897} (\bibinfo{year}{1998}).

\bibitem[{\citenamefont{Rathmann et~al.}(1998)}]{Rathmann:1998}
\bibinfo{author}{\bibfnamefont{F.}~\bibnamefont{Rathmann}}
  \bibnamefont{et~al.}, \bibinfo{journal}{Phys. Rev.}
  \textbf{\bibinfo{volume}{C58}}, \bibinfo{pages}{658} (\bibinfo{year}{1998}).

\bibitem[{\citenamefont{Ball et~al.}(2000)}]{Ball:2000}
\bibinfo{author}{\bibfnamefont{J.}~\bibnamefont{Ball}} \bibnamefont{et~al.},
  \bibinfo{journal}{CTU Reports} \textbf{\bibinfo{volume}{4}},
  \bibinfo{pages}{3} (\bibinfo{year}{2000}).

\bibitem[{\citenamefont{Arndt et~al.}(1994)\citenamefont{Arndt, Strakovskii,
  and Workman}}]{arndt94d}
\bibinfo{author}{\bibfnamefont{R.~A.} \bibnamefont{Arndt}},
  \bibinfo{author}{\bibfnamefont{I.~I.} \bibnamefont{Strakovskii}},
  \bibnamefont{and} \bibinfo{author}{\bibfnamefont{R.~L.}
  \bibnamefont{Workman}}, \bibinfo{journal}{Phys. Rev.}
  \textbf{\bibinfo{volume}{C50}}, \bibinfo{pages}{2731} (\bibinfo{year}{1994}).

\bibitem[{\citenamefont{Arndt et~al.}(1997)\citenamefont{Arndt, Oh, Strakovsky,
  Workman, and Dohrmann}}]{Arndt:1997if}
\bibinfo{author}{\bibfnamefont{R.~A.} \bibnamefont{Arndt}},
  \bibinfo{author}{\bibfnamefont{C.~H.} \bibnamefont{Oh}},
  \bibinfo{author}{\bibfnamefont{I.~I.} \bibnamefont{Strakovsky}},
  \bibinfo{author}{\bibfnamefont{R.~L.} \bibnamefont{Workman}},
  \bibnamefont{and} \bibinfo{author}{\bibfnamefont{F.}~\bibnamefont{Dohrmann}},
  \bibinfo{journal}{Phys. Rev.} \textbf{\bibinfo{volume}{C56}},
  \bibinfo{pages}{3005} (\bibinfo{year}{1997}), \eprint{nucl-th/9706003}.

\bibitem[{\citenamefont{Arndt et~al.}(2000)\citenamefont{Arndt, Strakovsky, and
  Workman}}]{Arndt:2000xc}
\bibinfo{author}{\bibfnamefont{R.~A.} \bibnamefont{Arndt}},
  \bibinfo{author}{\bibfnamefont{I.~I.} \bibnamefont{Strakovsky}},
  \bibnamefont{and} \bibinfo{author}{\bibfnamefont{R.~L.}
  \bibnamefont{Workman}}, \bibinfo{journal}{Phys. Rev. C}
  \textbf{\bibinfo{volume}{62}}, \bibinfo{pages}{34005} (\bibinfo{year}{2000}).

\bibitem[{\citenamefont{Ball et~al.}(1994)}]{ball94}
\bibinfo{author}{\bibfnamefont{J.}~\bibnamefont{Ball}} \bibnamefont{et~al.},
  \bibinfo{journal}{Phys. Lett. B} \textbf{\bibinfo{volume}{320}},
  \bibinfo{pages}{206} (\bibinfo{year}{1994}).

\bibitem[{\citenamefont{Gonzalez et~al.}(1987)\citenamefont{Gonzalez,
  {LaFrance}, and Lomon}}]{Gonzalez:1987gj}
\bibinfo{author}{\bibfnamefont{P.}~\bibnamefont{Gonzalez}},
  \bibinfo{author}{\bibfnamefont{P.}~\bibnamefont{{LaFrance}}},
  \bibnamefont{and} \bibinfo{author}{\bibfnamefont{E.~L.} \bibnamefont{Lomon}},
  \bibinfo{journal}{Phys. Rev.} \textbf{\bibinfo{volume}{D35}},
  \bibinfo{pages}{2142} (\bibinfo{year}{1987}).

\bibitem[{\citenamefont{Mulders et~al.}(1978)\citenamefont{Mulders, Aerts, and
  de~Swart}}]{Mulders:1978yi}
\bibinfo{author}{\bibfnamefont{P.~J.~G.} \bibnamefont{Mulders}},
  \bibinfo{author}{\bibfnamefont{A.~T.~M.} \bibnamefont{Aerts}},
  \bibnamefont{and} \bibinfo{author}{\bibfnamefont{J.~J.}
  \bibnamefont{de~Swart}}, \bibinfo{journal}{Phys. Rev. Lett.}
  \textbf{\bibinfo{volume}{40}}, \bibinfo{pages}{1543} (\bibinfo{year}{1978}).

\bibitem[{\citenamefont{Mulders et~al.}(1980)\citenamefont{Mulders, Aerts, and
  Swart}}]{Mulders:1980vx}
\bibinfo{author}{\bibfnamefont{P.~J.} \bibnamefont{Mulders}},
  \bibinfo{author}{\bibfnamefont{A.~T.} \bibnamefont{Aerts}}, \bibnamefont{and}
  \bibinfo{author}{\bibfnamefont{J.~J.~D.} \bibnamefont{Swart}},
  \bibinfo{journal}{Phys. Rev.} \textbf{\bibinfo{volume}{D21}},
  \bibinfo{pages}{2653} (\bibinfo{year}{1980}).

\bibitem[{\citenamefont{{LaFrance} and Lomon}(1986)}]{LaFrance:1986pc}
\bibinfo{author}{\bibfnamefont{P.}~\bibnamefont{{LaFrance}}} \bibnamefont{and}
  \bibinfo{author}{\bibfnamefont{E.~L.} \bibnamefont{Lomon}},
  \bibinfo{journal}{Phys. Rev.} \textbf{\bibinfo{volume}{D34}},
  \bibinfo{pages}{1341} (\bibinfo{year}{1986}).

\bibitem[{\citenamefont{{Albers} et~al.}(1997)}]{albers97}
\bibinfo{author}{\bibfnamefont{D.}~\bibnamefont{{Albers}}} \bibnamefont{et~al.}
  (\bibinfo{collaboration}{EDDA}), \bibinfo{journal}{Phys.\ Rev.\ Lett.}
  \textbf{\bibinfo{volume}{78}}, \bibinfo{pages}{1652} (\bibinfo{year}{1997}).

\bibitem[{\citenamefont{Altmeier et~al.}(2000)}]{Altmeier:2000}
\bibinfo{author}{\bibfnamefont{M.}~\bibnamefont{Altmeier}} \bibnamefont{et~al.}
  (\bibinfo{collaboration}{EDDA}), \bibinfo{journal}{Phys. Rev. Lett.}
  \textbf{\bibinfo{volume}{85}}, \bibinfo{pages}{1819} (\bibinfo{year}{2000}).

\bibitem[{\citenamefont{Bauer et~al.}(2003)}]{Bauer:2002zm}
\bibinfo{author}{\bibfnamefont{F.}~\bibnamefont{Bauer}} \bibnamefont{et~al.}
  (\bibinfo{collaboration}{EDDA}), \bibinfo{journal}{Phys. Rev. Lett.}
  \textbf{\bibinfo{volume}{90}}, \bibinfo{pages}{142301}
  (\bibinfo{year}{2003}).

\bibitem[{\citenamefont{Maier}(1997)}]{Maier:1997}
\bibinfo{author}{\bibfnamefont{R.}~\bibnamefont{Maier}},
  \bibinfo{journal}{Nucl.\ Instr.\ and Meth.} \textbf{\bibinfo{volume}{A390}},
  \bibinfo{pages}{1} (\bibinfo{year}{1997}).

\bibitem[{\citenamefont{Ackerstaff et~al.}(1993)}]{Ackerstaff:1993vy}
\bibinfo{author}{\bibfnamefont{K.}~\bibnamefont{Ackerstaff}}
  \bibnamefont{et~al.} (\bibinfo{collaboration}{EDDA}), \bibinfo{journal}{Nucl.
  Instrum. Meth.} \textbf{\bibinfo{volume}{A335}}, \bibinfo{pages}{113}
  (\bibinfo{year}{1993}).

\bibitem[{\citenamefont{Bisplinghoff et~al.}(1993)}]{bisplinghoff93}
\bibinfo{author}{\bibfnamefont{J.}~\bibnamefont{Bisplinghoff}}
  \bibnamefont{et~al.} (\bibinfo{collaboration}{EDDA}),
  \bibinfo{journal}{Nucl.\ Instr.\ and Meth.} \textbf{\bibinfo{volume}{A329}},
  \bibinfo{pages}{151} (\bibinfo{year}{1993}).

\bibitem[{\citenamefont{{Albers} et~al.}(1996)}]{albers96b}
\bibinfo{author}{\bibfnamefont{D.}~\bibnamefont{{Albers}}} \bibnamefont{et~al.}
  (\bibinfo{collaboration}{EDDA}), \bibinfo{journal}{Nucl.\ Instr.\ and Meth.}
  \textbf{\bibinfo{volume}{A 371}}, \bibinfo{pages}{388}
  (\bibinfo{year}{1996}).

\bibitem[{\citenamefont{Scheid}(1994)}]{scheid94}
\bibinfo{author}{\bibfnamefont{H.}~\bibnamefont{Scheid}},
  \bibinfo{type}{Dissertation}, \bibinfo{school}{Institut f\"ur Strahlen- und
  Kernphysik, Universit\"at Bonn} (\bibinfo{year}{1994}),
  \bibinfo{note}{\\\siehe: {\ttfon
  http://www.iskp.uni-bonn.de/edda/dipldiss.html}}.

\bibitem[{\citenamefont{Hinterberger and Prasuhn}(1989)}]{hin89}
\bibinfo{author}{\bibfnamefont{F.}~\bibnamefont{Hinterberger}}
  \bibnamefont{and} \bibinfo{author}{\bibfnamefont{D.}~\bibnamefont{Prasuhn}},
  \bibinfo{journal}{Nucl.\ Instr.\ and Meth.} \textbf{\bibinfo{volume}{A279}},
  \bibinfo{pages}{413} (\bibinfo{year}{1989}).

\bibitem[{\citenamefont{Lindlein}(1996)}]{lindlein96}
\bibinfo{author}{\bibfnamefont{J.}~\bibnamefont{Lindlein}},
  \bibinfo{type}{Diplomarbeit}, \bibinfo{school}{I.~Exp. Phys., Universit\"at
  Hamburg} (\bibinfo{year}{1996}), \bibinfo{note}{\\\siehe, {\ttfon
  http://kaa.desy.de/edda/papers/Diplomalist.inhalt.html}}.

\bibitem[{\citenamefont{Coffin}(1997)}]{COF97}
\bibinfo{author}{\bibfnamefont{J.~P.} \bibnamefont{Coffin}},
  \emph{\bibinfo{title}{Experimental Techniques in Nuclear Physics}}
  (\bibinfo{publisher}{de Gruyter}, \bibinfo{address}{Berlin},
  \bibinfo{year}{1997}), p. \bibinfo{pages}{309}.

\bibitem[{\citenamefont{Ackerstaff et~al.}(2002)}]{Ackerstaff:2002hj}
\bibinfo{author}{\bibfnamefont{K.}~\bibnamefont{Ackerstaff}}
  \bibnamefont{et~al.} (\bibinfo{collaboration}{EDDA}), \bibinfo{journal}{Nucl.
  Instrum. Meth.} \textbf{\bibinfo{volume}{A491}}, \bibinfo{pages}{492}
  (\bibinfo{year}{2002}).

\bibitem[{\citenamefont{Rohdje\ss{}
  et~al.}(2004{\natexlab{a}})}]{Rohdjess:2004b}
\bibinfo{author}{\bibfnamefont{H.}~\bibnamefont{Rohdje\ss{}}}
  \bibnamefont{et~al.} \bibnamefont{(EDDA)} , \eprint{nucl-ex/0403043}.

\bibitem[{\citenamefont{Bothe}(1995)}]{bothe95}
\bibinfo{author}{\bibfnamefont{H.-H.} \bibnamefont{Bothe}},
  \emph{\bibinfo{title}{Fuzzy Logic -- Einf\"uhrung in Theorie und
  Anwendungen}} (\bibinfo{publisher}{Springer, Heidelberg},
  \bibinfo{year}{1995}).

\bibitem[{\citenamefont{H{\"o}ppner et~al.}(1999)\citenamefont{H{\"o}ppner,
  Klawonn, Kruse, and Runkler}}]{hoppner1999}
\bibinfo{author}{\bibfnamefont{F.}~\bibnamefont{H{\"o}ppner}},
  \bibinfo{author}{\bibfnamefont{F.}~\bibnamefont{Klawonn}},
  \bibinfo{author}{\bibfnamefont{R.}~\bibnamefont{Kruse}}, \bibnamefont{and}
  \bibinfo{author}{\bibfnamefont{T.}~\bibnamefont{Runkler}},
  \emph{\bibinfo{title}{Fuzzy Cluster Analysis}} (\bibinfo{publisher}{Wiley,
  New York}, \bibinfo{year}{1999}).

\bibitem[{\citenamefont{Busch}(2001)}]{busch01}
\bibinfo{author}{\bibfnamefont{M.}~\bibnamefont{Busch}},
  \bibinfo{type}{{Dissertation}}, \bibinfo{school}{ISKP, Universit\"at Bonn}
  (\bibinfo{year}{2001}) \bibinfo{note}{\\\siehe: {\ttfon
  http://www.iskp.uni-bonn.de/edda/dipldiss.html}}.

\bibitem[{\citenamefont{Simon et~al.}(1993)}]{simon93}
\bibinfo{author}{\bibfnamefont{A.~J.} \bibnamefont{Simon}}
  \bibnamefont{et~al.}, \bibinfo{journal}{Phys. Rev. C}
  \textbf{\bibinfo{volume}{48}}, \bibinfo{pages}{662} (\bibinfo{year}{1993}).

\bibitem[{\citenamefont{Sternglass}(1957)}]{sternglass57}
\bibinfo{author}{\bibfnamefont{E.~J.} \bibnamefont{Sternglass}},
  \bibinfo{journal}{Phys. Rev.} \textbf{\bibinfo{volume}{108}},
  \bibinfo{pages}{1} (\bibinfo{year}{1957}).

\bibitem[{\citenamefont{Particle-Data-{Group, C. Caso et al.}}(1998)}]{pdg98}
\bibinfo{author}{\bibnamefont{Particle-Data-{Group, C. Caso et al.}}}
  (\bibinfo{collaboration}{Particle Data Group}), \bibinfo{journal}{Eur. Phys.
  J.} \textbf{\bibinfo{volume}{C3}}, \bibinfo{pages}{1} (\bibinfo{year}{1998}),
  \bibinfo{note}{see also {\ttfon http://pdg.lbl.gov}}.

\bibitem[{\citenamefont{Rosenbluth}(1950)}]{Rosenbluth1950}
\bibinfo{author}{\bibfnamefont{M.~N.} \bibnamefont{Rosenbluth}},
  \bibinfo{journal}{Phys. Rev.} \textbf{\bibinfo{volume}{79}},
  \bibinfo{pages}{615} (\bibinfo{year}{1950}).

\bibitem[{\citenamefont{Kallen}(1964)}]{kallen64E}
\bibinfo{author}{\bibfnamefont{G.}~\bibnamefont{Kallen}},
  \emph{\bibinfo{title}{{Elementary Particle Physics}}}
  (\bibinfo{publisher}{Addison-Wesley}, \bibinfo{year}{1964}).

\bibitem[{\citenamefont{H{\"u}skes}(1997)}]{hueskes97}
\bibinfo{author}{\bibfnamefont{T.}~\bibnamefont{H{\"u}skes}},
  \bibinfo{type}{Diplomarbeit}, \bibinfo{school}{{ISKP, Universit\"at Bonn}}
  (\bibinfo{year}{1997}), \bibinfo{note}{\\\siehe: {\ttfon
  http://www.iskp.uni-bonn.de/edda/dipldiss.html}}.

\bibitem[{\citenamefont{Nelson et~al.}(1985)\citenamefont{Nelson, Hirayama, and
  Rogers}}]{EGS4}
\bibinfo{author}{\bibfnamefont{W.~R.} \bibnamefont{Nelson}},
  \bibinfo{author}{\bibfnamefont{H.}~\bibnamefont{Hirayama}}, \bibnamefont{and}
  \bibinfo{author}{\bibfnamefont{D.~W.~O.} \bibnamefont{Rogers}},
  \emph{\bibinfo{title}{The EGS4 Code System}} (\bibinfo{year}{1985}),
  \bibinfo{note}{{SLAC-Report 265, URL:
  ``http://www.slac.stanford.edu/egs/''}}.

\bibitem[{\citenamefont{{Gro\ss{}-Hardt}}(2001)}]{gh01}
\bibinfo{author}{\bibfnamefont{R.}~\bibnamefont{{Gro\ss{}-Hardt}}},
  \bibinfo{type}{{Dissertation}}, \bibinfo{school}{{Institut f\"ur Strahlen-
  und Kernphysik, Universit\"at Bonn}} (\bibinfo{year}{2001}).

\bibitem[{\citenamefont{Rohdje\ss{}
  et~al.}(2004{\natexlab{b}})}]{Rohdjess:2004a}
\bibinfo{author}{\bibfnamefont{H.}~\bibnamefont{Rohdje\ss{}}}
  \bibnamefont{et~al.} \bibnamefont{(EDDA)}, \eprint{nucl-ex/0403044}.

\bibitem[{\citenamefont{Hinterberger and Prasuhn}(1992)}]{hin92}
\bibinfo{author}{\bibfnamefont{F.}~\bibnamefont{Hinterberger}}
  \bibnamefont{and} \bibinfo{author}{\bibfnamefont{D.}~\bibnamefont{Prasuhn}},
  \bibinfo{journal}{Nucl.\ Instr.\ and Meth.} \textbf{\bibinfo{volume}{A321}},
  \bibinfo{pages}{453} (\bibinfo{year}{1992}).

\bibitem[{\citenamefont{Engelhardt}(1998)}]{engelhardt98}
\bibinfo{author}{\bibfnamefont{H.~P.} \bibnamefont{Engelhardt}},
  \bibinfo{type}{Dissertation}, \bibinfo{school}{Institut f\"ur Strahlen- und
  Kernphysik, Universit\"at Bonn} (\bibinfo{year}{1998}),
  \bibinfo{note}{\\\siehe: {\ttfon
  http://www.iskp.uni-bonn.de/edda/dipldiss.html}}.

\bibitem[{Dat(2004)}]{DataAccess}
 \bibinfo{note}{Data can be accessed via http://kaa.desy.de and
  http://www.iskp.uni-bonn.de/gruppen/edda.}

\bibitem[{\citenamefont{Arndt et~al.}(1992)\citenamefont{Arndt, Roper, Workman,
  and McNaughton}}]{Arndt:1992kz}
\bibinfo{author}{\bibfnamefont{R.~A.} \bibnamefont{Arndt}},
  \bibinfo{author}{\bibfnamefont{L.~D.} \bibnamefont{Roper}},
  \bibinfo{author}{\bibfnamefont{R.~L.} \bibnamefont{Workman}},
  \bibnamefont{and} \bibinfo{author}{\bibfnamefont{M.~W.}
  \bibnamefont{McNaughton}}, \bibinfo{journal}{Phys. Rev.}
  \textbf{\bibinfo{volume}{D45}}, \bibinfo{pages}{3995} (\bibinfo{year}{1992}).

\bibitem[{\citenamefont{Arndt et~al.}(1985)\citenamefont{Arndt, Ford, and
  Roper}}]{Arndt:1985vj}
\bibinfo{author}{\bibfnamefont{R.~A.} \bibnamefont{Arndt}},
  \bibinfo{author}{\bibfnamefont{J.~M.} \bibnamefont{Ford}}, \bibnamefont{and}
  \bibinfo{author}{\bibfnamefont{L.~D.} \bibnamefont{Roper}},
  \bibinfo{journal}{Phys. Rev.} \textbf{\bibinfo{volume}{D32}},
  \bibinfo{pages}{1085} (\bibinfo{year}{1985}).

\bibitem[{\citenamefont{Arndt et~al.}(2004)\citenamefont{Arndt, Briscoe,
  Workman, and Strakovsky}}]{SAID}
\bibinfo{author}{\bibfnamefont{R.~A.} \bibnamefont{Arndt}},
  \bibinfo{author}{\bibfnamefont{W.~J.} \bibnamefont{Briscoe}},
  \bibinfo{author}{\bibfnamefont{R.~L.} \bibnamefont{Workman}},
  \bibnamefont{and} \bibinfo{author}{\bibfnamefont{I.~I.}
  \bibnamefont{Strakovsky}}, \emph{\bibinfo{title}{{SAID: {\bf S}cattering {\bf
  A}nalysis {\bf I}nteractive {\bf D}ialin}}} (\bibinfo{year}{2004}),
  \bibinfo{note}{http://gwdac.phys.gwu.edu/analysis/nn\_analysis}.

\bibitem[{\citenamefont{Jenkins et~al.}(1980)}]{jenkins80}
\bibinfo{author}{\bibfnamefont{K.~A.} \bibnamefont{Jenkins}}
  \bibnamefont{et~al.}, \bibinfo{journal}{Phys. Rev. D}
  \textbf{\bibinfo{volume}{21}}, \bibinfo{pages}{2445} (\bibinfo{year}{1980}).

\bibitem[{\citenamefont{Kammerud et~al.}(1971)}]{kammerud71}
\bibinfo{author}{\bibfnamefont{R.~C.} \bibnamefont{Kammerud}}
  \bibnamefont{et~al.}, \bibinfo{journal}{Phys. Rev. D}
  \textbf{\bibinfo{volume}{4}}, \bibinfo{pages}{1309} (\bibinfo{year}{1971}).

\bibitem[{\citenamefont{Williams et~al.}(1972)}]{williams72}
\bibinfo{author}{\bibfnamefont{D.~T.} \bibnamefont{Williams}}
  \bibnamefont{et~al.}, \bibinfo{journal}{Nuovo Cimento}
  \textbf{\bibinfo{volume}{8A}}, \bibinfo{pages}{447} (\bibinfo{year}{1972}).

\bibitem[{\citenamefont{Albrow et~al.}(1970)}]{albrow70}
\bibinfo{author}{\bibfnamefont{M.~G.} \bibnamefont{Albrow}}
  \bibnamefont{et~al.}, \bibinfo{journal}{Nucl.\ Phys.\ B}
  \textbf{\bibinfo{volume}{23}}, \bibinfo{pages}{445} (\bibinfo{year}{1970}).

\bibitem[{\citenamefont{Gar\c{con } et~al.}(1985)}]{garcon85}
\bibinfo{author}{\bibfnamefont{M.}~\bibnamefont{Gar\c{con }}}
  \bibnamefont{et~al.}, \bibinfo{journal}{Nucl. Phys. A}
  \textbf{\bibinfo{volume}{445}}, \bibinfo{pages}{669} (\bibinfo{year}{1985}).

\bibitem[{\citenamefont{Abe et~al.}(1975)}]{Abe:1975hs}
\bibinfo{author}{\bibfnamefont{K.}~\bibnamefont{Abe}} \bibnamefont{et~al.},
  \bibinfo{journal}{Phys. Rev.} \textbf{\bibinfo{volume}{D12}},
  \bibinfo{pages}{1} (\bibinfo{year}{1975}).

\bibitem[{\citenamefont{Simon et~al.}(1996)}]{simon96}
\bibinfo{author}{\bibfnamefont{A.~J.} \bibnamefont{Simon}}
  \bibnamefont{et~al.}, \bibinfo{journal}{Phys. Rev. C}
  \textbf{\bibinfo{volume}{53}}, \bibinfo{pages}{30} (\bibinfo{year}{1996}).

\bibitem[{\citenamefont{Hoffmann et~al.}(1988)}]{Hoffmann:1988jw}
\bibinfo{author}{\bibfnamefont{G.~W.} \bibnamefont{Hoffmann}}
  \bibnamefont{et~al.}, \bibinfo{journal}{Phys. Rev.}
  \textbf{\bibinfo{volume}{C37}}, \bibinfo{pages}{397} (\bibinfo{year}{1988}).

\bibitem[{\citenamefont{Ottewell et~al.}(1984)}]{Ottewell:1984hh}
\bibinfo{author}{\bibfnamefont{D.}~\bibnamefont{Ottewell}}
  \bibnamefont{et~al.}, \bibinfo{journal}{Nucl. Phys.}
  \textbf{\bibinfo{volume}{A412}}, \bibinfo{pages}{189} (\bibinfo{year}{1984}).

\bibitem[{\citenamefont{Dobrovolsky et~al.}(1983)}]{Dobrovolsky:1983wz}
\bibinfo{author}{\bibfnamefont{A.~v.} \bibnamefont{Dobrovolsky}}
  \bibnamefont{et~al.}, \bibinfo{journal}{Nucl. Phys.}
  \textbf{\bibinfo{volume}{B214}}, \bibinfo{pages}{1} (\bibinfo{year}{1983}).

\bibitem[{\citenamefont{l.~Barlett et~al.}(1983)}]{Barlett:1983ky}
\bibinfo{author}{\bibfnamefont{M.}~\bibnamefont{l.~Barlett}}
  \bibnamefont{et~al.}, \bibinfo{journal}{Phys. Rev.}
  \textbf{\bibinfo{volume}{C27}}, \bibinfo{pages}{682} (\bibinfo{year}{1983}).

\bibitem[{\citenamefont{Ryan et~al.}(1971)\citenamefont{Ryan, Kanofsky, Devlin,
  Mischke, and Shepard}}]{Ryan:1971rv}
\bibinfo{author}{\bibfnamefont{B.~A.} \bibnamefont{Ryan}},
  \bibinfo{author}{\bibfnamefont{A.}~\bibnamefont{Kanofsky}},
  \bibinfo{author}{\bibfnamefont{T.~J.} \bibnamefont{Devlin}},
  \bibinfo{author}{\bibfnamefont{R.~E.} \bibnamefont{Mischke}},
  \bibnamefont{and} \bibinfo{author}{\bibfnamefont{P.~F.}
  \bibnamefont{Shepard}}, \bibinfo{journal}{Phys. Rev.}
  \textbf{\bibinfo{volume}{D3}}, \bibinfo{pages}{1} (\bibinfo{year}{1971}).

\bibitem[{\citenamefont{Shimizu et~al.}(1982)}]{shimizu82}
\bibinfo{author}{\bibfnamefont{F.}~\bibnamefont{Shimizu}} \bibnamefont{et~al.},
  \bibinfo{journal}{Nucl. Phys.} \textbf{\bibinfo{volume}{A}},
  \bibinfo{pages}{445} (\bibinfo{year}{1982}).

\bibitem[{\citenamefont{Clyde}(1966)}]{Clyde:1966}
\bibinfo{author}{\bibfnamefont{A.~R.} \bibnamefont{Clyde}},
  \emph{\bibinfo{title}{{Lawrence Livermore Radiation Laboratory Report No.
  UCRL-16275}}} (\bibinfo{year}{1966}), \bibinfo{note}{unpublished}.

\bibitem[{\citenamefont{Ankenbrandt et~al.}(1968)}]{Ankenbrandt:1968}
\bibinfo{author}{\bibfnamefont{C.~M.} \bibnamefont{Ankenbrandt}}
  \bibnamefont{et~al.}, \bibinfo{journal}{Phys. Rev.}
  \textbf{\bibinfo{volume}{170}}, \bibinfo{pages}{1223} (\bibinfo{year}{1968}).

\bibitem[{\citenamefont{Rust et~al.}(1970)}]{Rust:1970wp}
\bibinfo{author}{\bibfnamefont{D.~R.} \bibnamefont{Rust}} \bibnamefont{et~al.},
  \bibinfo{journal}{Phys. Rev. Lett.} \textbf{\bibinfo{volume}{24}},
  \bibinfo{pages}{1361} (\bibinfo{year}{1970}).

\bibitem[{\citenamefont{Ambats et~al.}(1974)}]{ambats74}
\bibinfo{author}{\bibfnamefont{I.}~\bibnamefont{Ambats}} \bibnamefont{et~al.},
  \bibinfo{journal}{Phys.\ Rev.\ D} \textbf{\bibinfo{volume}{9}},
  \bibinfo{pages}{1179} (\bibinfo{year}{1974}).

\bibitem[{\citenamefont{Machleidt}(1989)}]{machleidt89}
\bibinfo{author}{\bibfnamefont{R.}~\bibnamefont{Machleidt}}, in
  \emph{\bibinfo{booktitle}{Adv. Nucl. Phys.}}, edited by
  \bibinfo{editor}{\bibfnamefont{J.~W.} \bibnamefont{Negele}} \bibnamefont{and}
  \bibinfo{editor}{\bibfnamefont{E.}~\bibnamefont{Vogt}}
  (\bibinfo{publisher}{Plenum Press, New York}, \bibinfo{year}{1989}),
  vol.~\bibinfo{volume}{19}, chap.~\bibinfo{chapter}{2}, pp.
  \bibinfo{pages}{189--376}.

\bibitem[{\citenamefont{van Kolck}(1999)}]{vanKolck:1999mw}
\bibinfo{author}{\bibfnamefont{U.}~\bibnamefont{van Kolck}},
  \bibinfo{journal}{Prog. Part. Nucl. Phys.} \textbf{\bibinfo{volume}{43}},
  \bibinfo{pages}{337} (\bibinfo{year}{1999}).

\bibitem[{\citenamefont{Epelbaum et~al.}(2000)\citenamefont{Epelbaum,
  Gl\"ockle, and Meissner}}]{Epelbaum:1999dj}
\bibinfo{author}{\bibfnamefont{E.}~\bibnamefont{Epelbaum}},
  \bibinfo{author}{\bibfnamefont{W.}~\bibnamefont{Gl\"ockle}},
  \bibnamefont{and} \bibinfo{author}{\bibfnamefont{U.-G.}
  \bibnamefont{Meissner}}, \bibinfo{journal}{Nucl. Phys.}
  \textbf{\bibinfo{volume}{A671}}, \bibinfo{pages}{295} (\bibinfo{year}{2000}),
  \eprint{nucl-th/9910064}.

\bibitem[{\citenamefont{Entem and Machleidt}(2003)}]{Entem:2003ft}
\bibinfo{author}{\bibfnamefont{D.~R.} \bibnamefont{Entem}} \bibnamefont{and}
  \bibinfo{author}{\bibfnamefont{R.}~\bibnamefont{Machleidt}},
  \bibinfo{journal}{Phys. Rev.} \textbf{\bibinfo{volume}{C68}},
  \bibinfo{pages}{041001} (\bibinfo{year}{2003}), \eprint{nucl-th/0304018}.

\bibitem[{\citenamefont{Elster et~al.}(1988{\natexlab{a}})\citenamefont{Elster,
  Holinde, Sch{\"u}tte, and Machleidt}}]{Elster:1988zu}
\bibinfo{author}{\bibfnamefont{C.}~\bibnamefont{Elster}},
  \bibinfo{author}{\bibfnamefont{K.}~\bibnamefont{Holinde}},
  \bibinfo{author}{\bibfnamefont{D.}~\bibnamefont{Sch{\"u}tte}},
  \bibnamefont{and}
  \bibinfo{author}{\bibfnamefont{R.}~\bibnamefont{Machleidt}},
  \bibinfo{journal}{Phys. Rev.} \textbf{\bibinfo{volume}{C38}},
  \bibinfo{pages}{1828} (\bibinfo{year}{1988}{\natexlab{a}}).

\bibitem[{\citenamefont{Elster et~al.}(1988{\natexlab{b}})\citenamefont{Elster,
  Ferchl{\"a}nder, Holinde, Sch{\"u}tte, and Machleidt}}]{Elster:1988pp}
\bibinfo{author}{\bibfnamefont{C.}~\bibnamefont{Elster}},
  \bibinfo{author}{\bibfnamefont{W.}~\bibnamefont{Ferchl{\"a}nder}},
  \bibinfo{author}{\bibfnamefont{K.}~\bibnamefont{Holinde}},
  \bibinfo{author}{\bibfnamefont{D.}~\bibnamefont{Sch{\"u}tte}},
  \bibnamefont{and}
  \bibinfo{author}{\bibfnamefont{R.}~\bibnamefont{Machleidt}},
  \bibinfo{journal}{Phys. Rev.} \textbf{\bibinfo{volume}{C37}},
  \bibinfo{pages}{1647} (\bibinfo{year}{1988}{\natexlab{b}}).

\bibitem[{\citenamefont{Eyser et~al.}(2003)\citenamefont{Eyser, Machleidt, and
  Scobel}}]{Eyser:2003zw}
\bibinfo{author}{\bibfnamefont{K.~O.} \bibnamefont{Eyser}},
  \bibinfo{author}{\bibfnamefont{R.}~\bibnamefont{Machleidt}},
  \bibnamefont{and} \bibinfo{author}{\bibfnamefont{W.}~\bibnamefont{Scobel}}
  (\bibinfo{year}{2003}), \eprint{nucl-th/0311002}.

\bibitem[{\citenamefont{Elster}(2003)}]{Elster:2004}
\bibinfo{author}{\bibfnamefont{C.}~\bibnamefont{Elster}},
  \bibinfo{howpublished}{Ohio University, Athens, Ohio} (\bibinfo{year}{2003}),
  \bibinfo{note}{private communication}.

\bibitem[{\citenamefont{Brodsky and Farrar}(1973)}]{Brodsky:1973kr}
\bibinfo{author}{\bibfnamefont{S.~J.} \bibnamefont{Brodsky}} \bibnamefont{and}
  \bibinfo{author}{\bibfnamefont{G.~R.} \bibnamefont{Farrar}},
  \bibinfo{journal}{Phys. Rev. Lett.} \textbf{\bibinfo{volume}{31}},
  \bibinfo{pages}{1153} (\bibinfo{year}{1973}).

\bibitem[{\citenamefont{Auer et~al.}(1978)}]{Auer:1978aa}
\bibinfo{author}{\bibfnamefont{I.~P.} \bibnamefont{Auer}} \bibnamefont{et~al.},
  \bibinfo{journal}{Phys. Rev. Lett.} \textbf{\bibinfo{volume}{41}},
  \bibinfo{pages}{354} (\bibinfo{year}{1978}).

\bibitem[{\citenamefont{Biegert et~al.}(1978)}]{bieler78}
\bibinfo{author}{\bibfnamefont{E.~K.} \bibnamefont{Biegert}}
  \bibnamefont{et~al.}, \bibinfo{journal}{Phys. Lett.}
  \textbf{\bibinfo{volume}{73B}}, \bibinfo{pages}{235} (\bibinfo{year}{1978}).

\bibitem[{\citenamefont{Bugg}(1984)}]{bug84}
\bibinfo{author}{\bibfnamefont{D.~V.} \bibnamefont{Bugg}}, \bibinfo{journal}{J.
  Phys. G} \textbf{\bibinfo{volume}{10}}, \bibinfo{pages}{717}
  (\bibinfo{year}{1984}).

\bibitem[{\citenamefont{Gonzalez and Lomon}(1986)}]{Gonzalez:1986pd}
\bibinfo{author}{\bibfnamefont{P.}~\bibnamefont{Gonzalez}} \bibnamefont{and}
  \bibinfo{author}{\bibfnamefont{E.~L.} \bibnamefont{Lomon}},
  \bibinfo{journal}{Phys. Rev.} \textbf{\bibinfo{volume}{D34}},
  \bibinfo{pages}{1351} (\bibinfo{year}{1986}).

\bibitem[{\citenamefont{{LaFrance} et~al.}(1993)\citenamefont{{LaFrance},
  Lomon, and Aw}}]{LaFrance:1993aq}
\bibinfo{author}{\bibfnamefont{P.}~\bibnamefont{{LaFrance}}},
  \bibinfo{author}{\bibfnamefont{E.~L.} \bibnamefont{Lomon}}, \bibnamefont{and}
  \bibinfo{author}{\bibfnamefont{M.}~\bibnamefont{Aw}} (\bibinfo{year}{1993}),
  \eprint{nucl-th/9306026}.

\bibitem[{\citenamefont{Ueda}(1982)}]{Ueda:1982fz}
\bibinfo{author}{\bibfnamefont{T.}~\bibnamefont{Ueda}}, \bibinfo{journal}{Phys.
  Lett.} \textbf{\bibinfo{volume}{B119}}, \bibinfo{pages}{281}
  (\bibinfo{year}{1982}).

\end{thebibliography}
\end {document}